\documentclass[twocolumn,amsfonts,amssymb,amsmath,superscriptaddress]{revtex4-1}

\usepackage{graphicx} %
\usepackage{float} %
\usepackage{soul} %
\graphicspath{{figs/}} %
\usepackage[colorlinks = true, %
linkcolor = blue,
urlcolor  = blue,
citecolor = blue,
anchorcolor = blue]{hyperref}
\usepackage[dvipsnames]{xcolor} %

\usepackage[T1]{fontenc}
\usepackage[utf8]{inputenc}
\usepackage[normalem]{ulem}

\usepackage{braket} %
\usepackage{makecell} %

\begin{document}
\date{May 14, 2025}
\title{Emergent chirality and enantiomeric selectivity in layered NbOX$_2$ crystals}

\author{Martin Gutierrez-Amigo}
\thanks{Corresponding author - Martin Gutierrez-Amigo \\
    martin.gutierrezamigo@aalto.fi
}

\affiliation{Department of Applied Physics, Aalto University School of Science, FI-00076 Aalto, Finland}

\author{Claudia Felser}
\affiliation{Max Planck Institute for Chemical Physics of Solids, 01187 Dresden, Germany}

\author{Ion Errea}
\affiliation{Centro de Física de Materiales (CFM-MPC), CSIC-UPV/EHU, 20018 Donostia/San Sebasti\'an, Spain}
\affiliation{Fisika Aplikatua Saila, Gipuzkoako Ingeniaritza Eskola, University of the Basque Country (UPV/EHU), 20018 Donostia/San Sebasti\'an, Spain}
\affiliation{Donostia International Physics Center (DIPC), 20018 Donostia/San Sebasti\'an, Spain}

\author{Maia G. Vergniory}
\affiliation{Département de Physique et Institut Quantique, Université de Sherbrooke, Sherbrooke, J1K 2R1 Québec, Canada.}
\affiliation{Donostia International Physics Center (DIPC), 20018 Donostia/San Sebasti\'an, Spain}
\affiliation{Regroupement Qu\'eb\'ecois sur les Mat\'eriaux de Pointe (RQMP), Quebec H3T 3J7, Canada}

\begin{abstract}
    The spontaneous emergence of chirality in crystalline solids has profound implications for electronic, optical, and topological properties, making the control of chiral phases a central challenge in materials design.
    Here, we investigate the structural and electronic properties of a new family of layered compounds, $\mathrm{NbOX_2}$, and explore the connection between their achiral $I m m m$ phase and chiral $C 2$ phase.
    Through first-principles calculations, we identify an intermediate achiral $C 2/m$ phase that bridges the high- and low-symmetry phases within a three-dimensional order parameter space.
    By analyzing the Born-Oppenheimer energy surfaces, we find that the shallow energy minima of the $C2/m$ phase suggest it may be stabilized either by external factors such as pressure, as demonstrated here, or by ionic quantum or thermal fluctuations and the resulting lattice anharmonicity.
    Additionally, we show how an external electric field, by breaking the necessary symmetries, biases the system toward a preferred chirality by lifting the energy degeneracy between the two enantiomers.
    This, combined with the small energy barrier between the enantiomers in the $C 2$ phase, enables handedness control and allows us to propose a mechanism for selective handedness stabilization by leveraging electric fields and pressure or temperature-dependent anharmonic effects.
    Our findings establish a framework for understanding chirality emergence in layered materials and offer a pathway for designing systems with tunable enantiomeric populations.
\end{abstract}

\maketitle
\newpage

Chirality is the asymmetry of a particular object which cannot be superimposed over it's specular image.
These mirror-image counterparts, called enantiomers, may exhibit dramatically different properties despite being related by a simple mirror operation.
The concept of chirality has been historically central to organic chemistry, where enantiomers of a molecule can differ significantly, exhibiting distinct smells or having varying effects as drugs.
Beyond organic chemistry, in inorganic systems, chirality underpins a diverse range of phenomena, including circular dichroism, circular photogalvanic effects \cite{dejuan_2017}, chiral-induced spin selectivity \cite{naaman_2012,evers_2022}, chiral phonons \cite{zhang_2015,zhu_2018}, and the emergence of novel topological states of matter \cite{chang_2018}.
These effects hold promise for applications in optoelectronic devices \cite{dor_2013,long_2020}, catalysts for enantioselective reactions \cite{song_2002,switzer_2003,yoon_2003}, and spintronics \cite{naaman_2015,naaman_2019,yang_2021}.
A particularly desirable feature of chiral materials is their ability to achieve chiral selectivity, the controlled spontaneous breaking of improper symmetries to favor one enantiomer over the other, as such control could pave the way for new functionalities in technology and materials science.

A promising strategy to achieve chiral selectivity is to identify materials with achiral phases energetically close to their chiral ground state, enabling chirality switching through an intermediate achiral phase.
In this context, the family of ferroelectric van der Waals materials $\mathrm{NbOX_2}$ (X = I, Br, Cl) has emerged as a particularly exciting candidate.
These materials exhibit a low-energy barrier for switching ferroelectric (FE) ordering, facilitating efficient polarization manipulation \cite{jia_2019, chen_2021c, abdelwahab_2022}.
Unlike many 2D ferroelectric materials, their ferroelectric properties are layer-independent, extending from monolayers to bulk structures \cite{jia_2019, chen_2021c}, making them ideal for nanoscale devices.
In addition, $\mathrm{NbOX_2}$ compounds display exceptional nonlinear optical properties, including record-setting second-harmonic generation efficiencies that scale with layer thickness \cite{abdelwahab_2022, guo_2023}, making them promising for polarization-sensitive optoelectronic devices such as photodetectors and sensors \cite{fang_2021, huang_2024}.
This combination of properties positions $\mathrm{NbOX_2}$ compounds as an ideal platform for exploring the interplay between chirality, ferroelectricity, and external stimuli.

\begin{figure*}[]
    \centering
    \includegraphics[width=\textwidth]{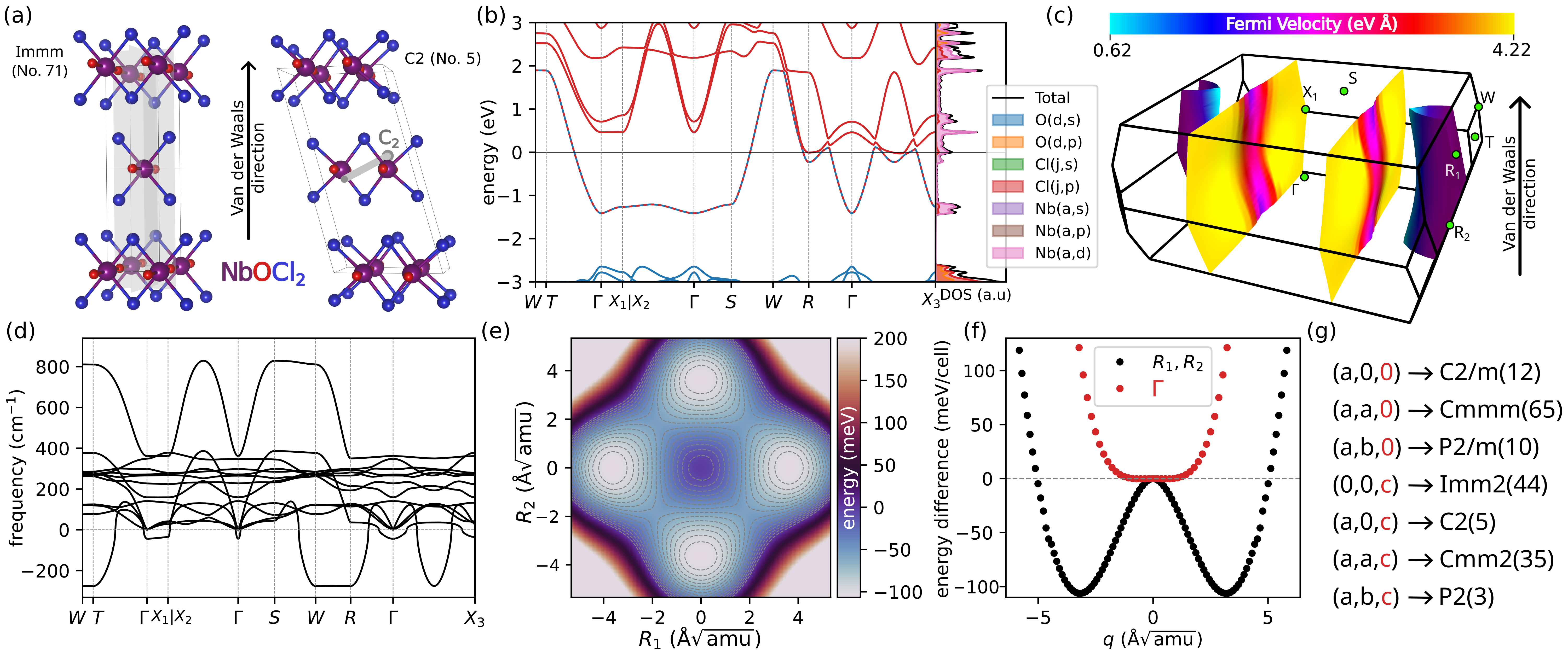}
    \caption{
        \textbf{$\mathbf{NbOCl_2}$ high-symmetry phase SG $\mathbf{Im m m}$  (No. 71).}
        \textbf{a.} $\mathrm{NbOCl_2}$ high-symmetry structure (left) compared to the $\mathrm{NbOCl_2}$ low-symmetry phase (right).
        \textbf{b.} Electronic band structure of the high-symmetry phase of $\mathrm{NbOCl_2}$, with atomic orbital projections. Labels indicate the Wyckoff position (first letter) and orbital character (second letter) of each contribution.
        Valence and conduction bands are shown in blue and red, respectively; the central band is rendered in a mixed color to reflect the system’s odd electron count.
        \textbf{c.} Fermi surface of the high-symmetry phase with the reciprocal space high-symmetry points over the first Brillouin zone. The black line indicates the Van der Waals direction in reciprocal space.
        \textbf{d.} Calculated harmonic phonon dispersion for the high-symmetry phase.
        \textbf{e.} Born-Oppenheimer energy landscape for the $\left\{ R_1,R_2 \right\} $ surface of the order parameter space, with the origin corresponding to the high-symmetry phase.
        Four energy minimas appear in the $(\pm R_{1},0)$ and $(0,\pm R_{2})$ directions.
        \textbf{f.} Computed Born-Oppenheimer energies as the high-symmetry structure is distorted either parallel to the $\Gamma_{3}^{-}$ or ($R_{2}^{-}$,$R_{1}^{-}$) modes.
        \textbf{g.} All possible space groups spanning from the three-dimensional order parameter $(R_{1},R_{2},\Gamma)$.
        In this context, $(a, b, c)$ represent non-zero real numbers corresponding to the amplitudes of the order parameters.
    }
    \label{fig:1}
\end{figure*}

One particularly interesting aspect of the $\mathrm{NbOX_2}$ family lies in its structural diversity.
The $\mathrm{NbOX_2}$ materials have been experimentally reported to crystallize in space group (SG) $C2$ (No. 5), a chiral structure \cite{rijnsdorp_1978,seifert_1981,beck_2006}.
However, $\mathrm{NbOCl_2}$ has also been reported in the achiral and higher-symmetry space group $Immm$ (No. 71) \cite{hillebrecht_1997}.
The fact that the same compound has been reported in both a chiral and an achiral group (of much higher symmetry), opens the possibility of having a phase transition mediating between the two crystal structures.
A structural comparison between the reported $Immm$ and $C2$ phases of $\mathrm{NbOCl_2}$ reveals subtle symmetry-breaking distortions in the chiral phase (Fig. \ref{fig:1}(a)), indicating that a charge-density wave (CDW) could mediate the transition.
This mechanism supports both the possibility of chiral selectivity and the design of materials with tunable chirality via control of intermediate CDW states.

In this work, we use first-principles calculations to investigate the structural and electronic properties of $\mathrm{NbOX_2}$ compounds, focusing on the connection between the reported $Immm$ and $C2$ phases of $\mathrm{NbOCl_2}$.
By identifying and characterizing the CDWs that mediate this transition, we aim to establish a method for achieving chiral selectivity in these materials.
To this end, we analyze the Born-Oppenheimer energy surfaces (BOES) associated with the relevant phonon modes and identify the $C 2/m$ phase as a low-energy achiral intermediate structure, which can potentially be stabilized through anharmonic effects, external pressure or photoexcitation.
We further characterize the electronic structure of all three phases, with the $C 2$ phase being an obstructed atomic limit (OAL) phase that hosts topologically non-trivial surface states under specific cleavage conditions.
Additionally, we investigate the small energy barrier between enantiomers of the $C 2$ phase, demonstrating how an external electric field can lift the energy degeneracy and bias the system toward a specific handedness.
Lastly, we propose a practical mechanism for achieving chirality selectivity by combining electric fields and thermal fluctuations, offering a pathway for designing chiral materials with controllable enantiomeric populations.
While the main manuscript focuses on \(\mathrm{NbOCl_2}\) as a representative example of the \(\mathrm{NbOX_2}\) family, the same systematic analysis has been applied to \(\mathrm{NbOBr_2}\) and \(\mathrm{NbOI_2}\), yielding the same qualitative results (see supplementary material \cite{supplementary}).

The reported structure of $\mathrm{NbOCl_2}$ in space group $I m m m$ (No. 71) \cite{hillebrecht_1997} exhibits disorder in the chlorine atoms.
Therefore, to enable density functional theory (DFT) calculations, we first constructed an ordered arrangement of chlorine atoms to compute both the phonon and electronic band structures.
The electronic bands shown in Fig. \ref{fig:1}(b) show that the $I m m m$ is metallic (forced by the odd number of electrons), with a very low dispersion parallel to the Van der Waals (VdW) direction, showcasing the layered nature of the compound (see Fig. \ref{fig:1}(b,c)).
Structures with an odd number of electrons are inherently prone to instabilities arising from spin imbalance, often resulting in a reconstruction of the electronic states.
In this case, the phonon spectrum displays pronounced instabilities, as shown in Fig. \ref{fig:1}(d), highlighting the structural susceptibility of the compound and providing a natural explanation for its tendency toward disorder.
Additionally, the Fermi surface displayed in Fig. \ref{fig:1}(c) shows the anticipated lack of dispersion in the VdW plane as well as two flat Fermi sheets akin to the ones one would expect from a one-dimensional system.
This Fermi surface is dominated by the Niobium $d$ orbitals, which are predisposed to reconstruct and stabilize the system by forming a lower-energy, insulating state.

The structural instability implied by the phonon spectrum leads to a multidimensional order parameter, making it nontrivial to identify the modes responsible for the CDW.
However, group-theoretical constraints and the fact that the transition leads to the chiral $C2$ structure (No.~5) \cite{seifert_1981,guo_2023} restrict the possibilities to a small number of symmetry-allowed mode combinations (see supplementary material \cite{supplementary}).
The order parameter connecting the space groups $Immm$ and $C2$ can thus be reduced to a three-dimensional vector, $(R_1, R_2, \Gamma)$, where the subscripts of $R$ represent the two symmetry-equivalent directions (star vectors) associated with the unstable mode $R_{2}^{-}$.
To gain physical insight into the energy gain associated with different directions in this order parameter space, we compute the BOES along these three directions, as shown in Fig. \ref{fig:1}(f).
The energy landscape in the $(R_1, R_2)$ plane exhibits a characteristic "Mexican hat" shape, with a minimum energy reduction of approximately $\sim  100\ \mathrm{meV/cell}$.
In contrast, the BOES along the $\Gamma$ direction is nearly flat, even for large displacements.
These results suggest that while the $\Gamma_{3}^{-}$ mode could stabilize under anharmonic effects, optical excitation \cite{rodriguez-vega_2022}, or applied pressure, the $R_{2}^{-}$ modes are unlikely to stabilize due to the significant depth of the energy well.
Instead, these modes are likely responsible for the observed disorder in the $Immm$ phase \cite{hillebrecht_1997}.

Building on these insights, we first focus on resolving the condensation of the $R_{2}^{-}$ modes independently of the $\Gamma_{3}^{-}$ mode, which, according to the BOES, could stabilize under specific conditions.
To this end, we sample the Born-Oppenheimer energy surface across the whole two-dimensional plane spanned by $(R_1, R_2, 0)$ in the order parameter space (see Fig. \ref{fig:1}(e)).
Our analysis reveals four equivalent energy minima on the BOES, each corresponding to the independent condensation of either $R_1$ or $R_2$.
This outcome aligns with our expectations, as any combination of the $R_{2}^{-}$ modes would result in structures with 16 atoms per primitive cell, failing to satisfy the requirement of having eight atoms in the final structure as reported.
The structures that minimize the BOES within the $(R_1, R_2, 0)$ plane belong to the $C2/m$ space group (No. 12), which is achiral.
This group has also been reported in the literature \cite{jia_2019,fu_2024,zhang_2024} and serves as a parent phase from which both enantiomers of the $C2$ structure can emerge.

\begin{figure}[]
    \centering
    \includegraphics[width=\linewidth]{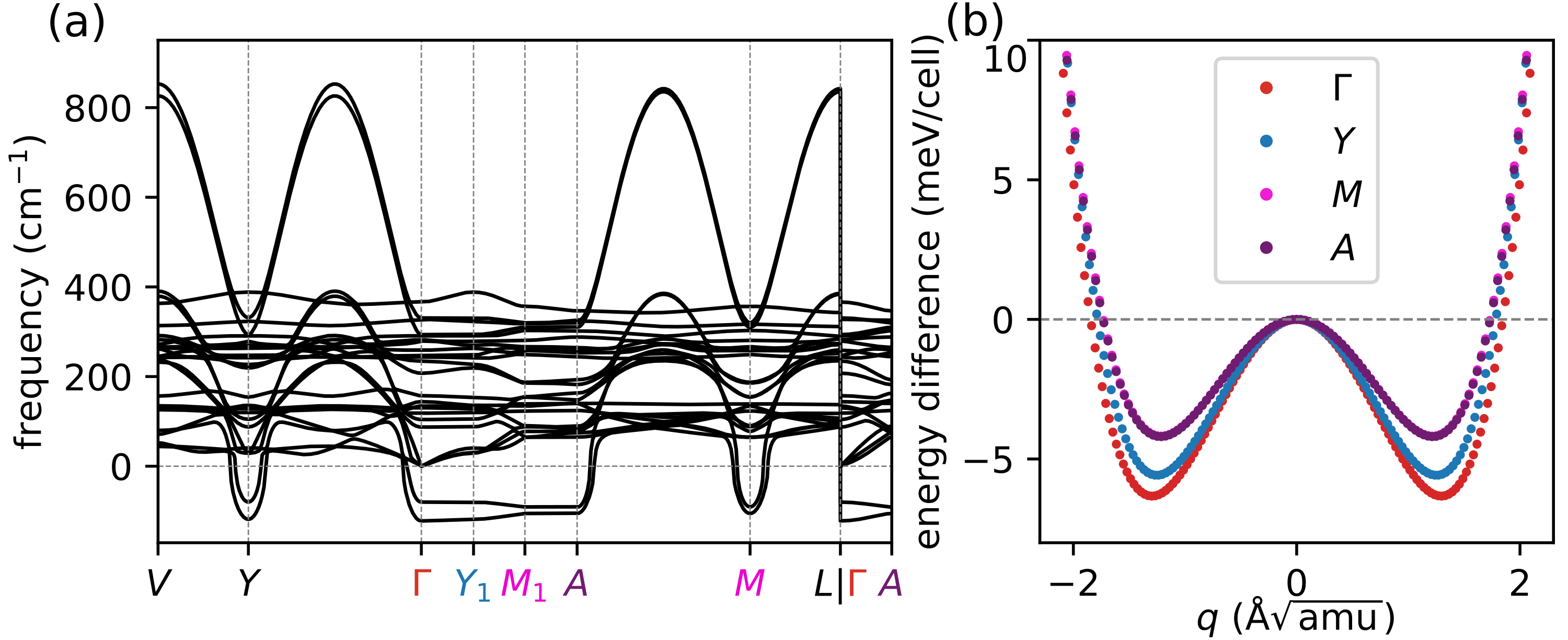}
    \caption{
        \textbf{$\mathbf{NbOCl_2}$ intermediate-symmetry phase SG $\mathbf{C 2/m}$  (No. 12).}
        \textbf{a.} Phonon spectrum of the intermediate insulating phase. The negative phonons at the  $\Gamma,Y,M$ and $A$ points imply that the structure is not thermodynamically stable under the harmonic approximation.
        \textbf{b.} BOES as the structure distorts according to each of the most unstable phonons at each high-symmetry point, where the $\Gamma$ mode leads to the lowest BO energy.
    }
    \label{fig:2}
\end{figure}

After relaxing the lattice parameters and ionic positions for the $C2/m$ phase, we recomputed the band structure and phonon spectrum.
As anticipated, the CDW driven by the $R_{2}^{-}$ modes leads to a metal-insulator phase transition, with the low-symmetry phase featuring a flat band near the Fermi level formed by recombined niobium $d$ orbitals.
Interestingly, this band corresponds to an OAL \cite{ma_2023,cano_2022,xu_2024a}, indicating that it is related to a single Wannier center that does not lie over the ionic positions.
This implies that the surface is expected to host topologically non-trivial surface states, provided the cleavage surface intersects the charge center, which in this case is at the \( 2a \) Wyckoff position.
As a result, the underlying mechanism through which the CDW lowers the free energy is likely the enhancement of electronic localization across the \( 2a \) Wyckoff position.
This increased localization, in turn, accounts for the band’s flatness and its lack of dispersion, even in directions perpendicular to the van der Waals direction.
Regarding stability, the phonon spectrum in Fig. \ref{fig:2}(a) still shows instabilities, as well as the characteristic lack of dispersion along the van der Waals direction (see Figs. \ref{fig:2}(a) and S5).
To further explore the system's energy minimization, we compute the BOES across the most unstable phonon modes at high-symmetry points $\Gamma_{1}^{-}$, $Y_1^{-}$, $M_2^{+}$, and $A_2^{+}$.
As shown in Fig. \ref{fig:2}(b), the BOES for all modes exhibit similar features, with the $\Gamma_{1}^{-}$ mode leading to the lowest energy.
Given the profile of the BOES and the fact that the $\Gamma$ mode is the only compatible phonon with the low-symmetry $C2$ phase, we identify the $\Gamma_{1}^{-}$ mode as the primary instability in the system.
Notably, the symmetry properties of $\Gamma_{1}^{-}$ correspond precisely to those of the $\Gamma_{3}^{-}$ mode in the high-symmetry $Immm$ phase when the representation is subduced to the $C2/m$ space group.
Unlike the flat BOES of $\Gamma_{3}^{-}$ in the $Immm$ phase, the BOES of $\Gamma_{1}^{-}$ in the $C2/m$ phase now resembles a Mexican hat shape, albeit with much smaller minima $\sim 7\ \mathrm{meV/cell}$ than the $R_{2}^{-}$ modes in the $I m m m$ phase.

Due to the shallowness of the potential wells, the Born-Oppenheimer energy surfaces for the various unstable modes shown in Fig. \ref{fig:2}(b) suggest that the achiral \( C2/m \) phase could potentially be stabilized, serving as the proposed parent space group from which both enantiomers could emerge.
We test this assumption by computing the evolution of the unstable \(\Gamma_{1}^{-}\) phonon with respect to pressure and find that the intermediate \(C2/m\) phase is predicted to be stabilized around 81 kbar (see supplementary material \cite{supplementary}).
Notably, the critical pressure decreases systematically across the NbOX$_2$ series from Cl to I.
In addition, a recent study \cite{cardenas-gamboa_2025} shows that optical excitation can also enhance the metastability of intermediate phase.
Beyond using external perturbations such as pressure or photoexcitation, typical thermal energies, where 100 K corresponds to approximately \( \sim 8\ \mathrm{meV} \), indicate that quantum or thermal ionic fluctuations and the resulting anharmonicity could also play a significant role in stabilizing the \( C2/m \) phase.
These findings are consistent with previous studies that have demonstrated the role of anharmonicity in stabilizing phases with shallow energy wells at finite temperatures or under different external conditions \cite{errea_2016, errea_2020, fumega_2023, gutierrez-amigo_2024, gutierrez-amigo_2024b}.

\begin{figure}[]
    \centering
    \includegraphics[width=\linewidth]{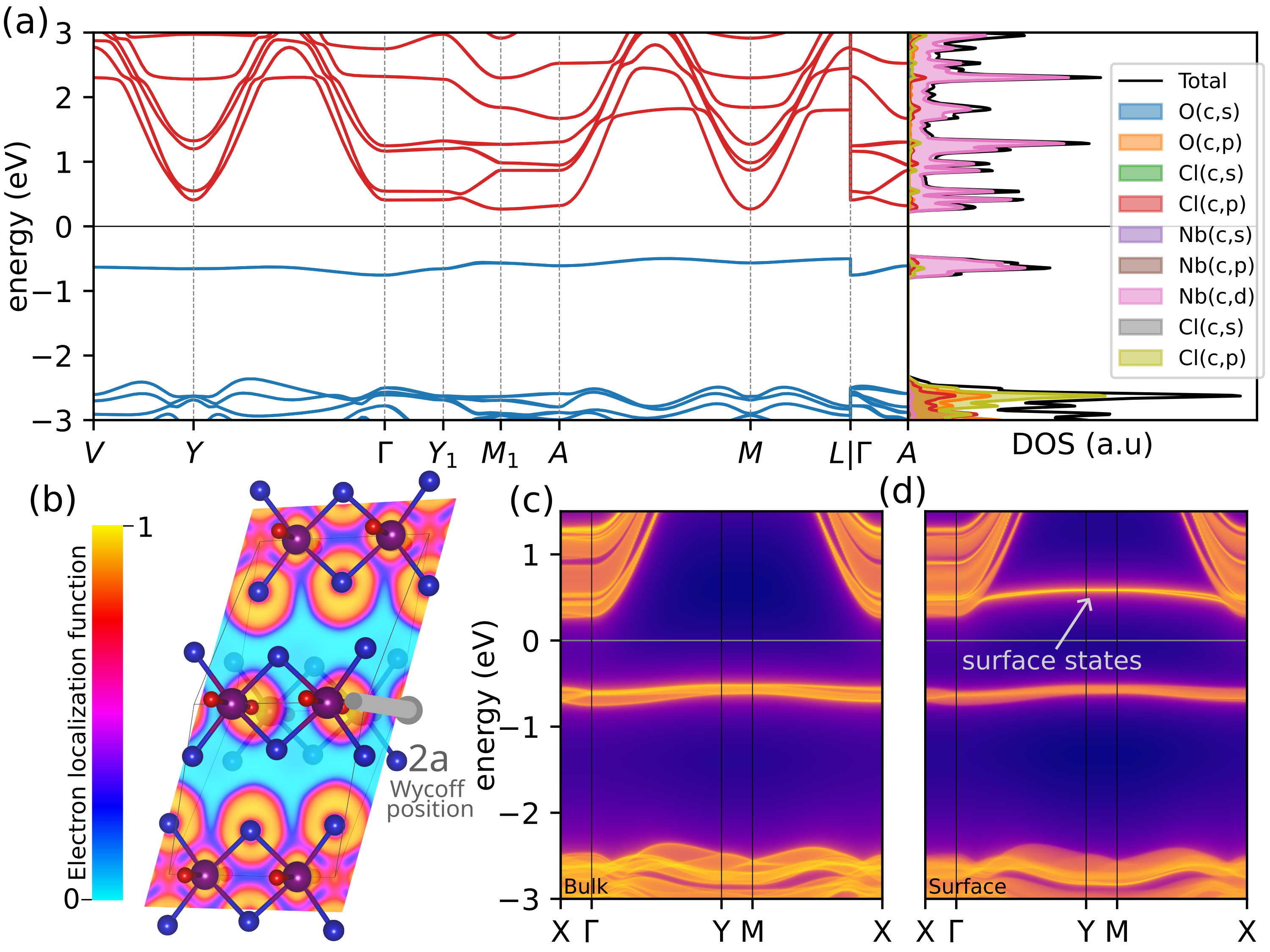}
    \caption{
        \textbf{$\mathbf{NbOCl_2}$ low-symmetry phase SG $\mathbf{C 2}$ (No. 5).}
        \textbf{a.} Band structure of the insulating low-symmetry phase. The isolated valence formed by Niobium $d$ orbitals at around $0.5$ eV constitutes an obstructed atomic limit centered at the Wyckoff position $2a$ indicated in inset (b).
        \textbf{b.} Electron localization function (ELF) of the low-symmetry phase showcasing low electronic localization in Miller planes (110) perpendicular to the Van der Waals stacking.
        \textbf{c, d.} Surface state ARPES simulation for the $(112)$ surface. In \textbf{c} only the bulk states are represented, while in \textbf{d} also the surface states are included.
    }
    \label{fig:3}
\end{figure}

As initially proposed, the structures that minimize the BOES spanned by the $\Gamma_{1}^{-}$ mode belong to the chiral space group $C2$ (No. 5).
Akin to the previous procedure, we relax both the lattice parameters and the ionic positions, resulting in a structure that closely matches those reported in the literature \cite{seifert_1981}.
In contrast to the other phases, and in agreement with experimental evidence \cite{guo_2023,liu_2023,fu_2024}, the phonon spectrum of the $C2$ phase (see Fig. S6) exhibits no negative phonon frequencies, indicating that it is an stable phase.
Similar to the $C2/m$ phase, the $C2$ phase is also insulating, with an isolated flat band near the Fermi level composed of Niobium $d$ orbitals (see Fig. \ref{fig:3}(a)).
This flat band might be of interest, as its limited dispersion may enhance the role of electron–electron interactions, favoring correlated electronic behavior.
Furthermore, these bands also correspond to an OAL, and consequently, give rise to non-trivial surface states, as shown in Fig. \ref{fig:3}(c,d), when the surface intersects the $2a$ Wyckoff position (see supplementary material \cite{supplementary} for further details).
To determine whether the natural cleavage surface of the $C2$ phase corresponds to one of these surfaces, we compute the electron localization function (ELF).
As shown in Fig. \ref{fig:3}(b), the ELF exhibits a constant low value across the $110$ surface, which is perpendicular to the Van der Waals stacking.
This low electron density suggests a lack of strong bonding and defines the natural cleavage surface.
Consequently, the $110$ surface will not host surface states (see Fig. S4(f)) as it does not intersect any of the Wannier charge centers.

\begin{figure*}[]
    \centering
    \includegraphics[width=\textwidth]{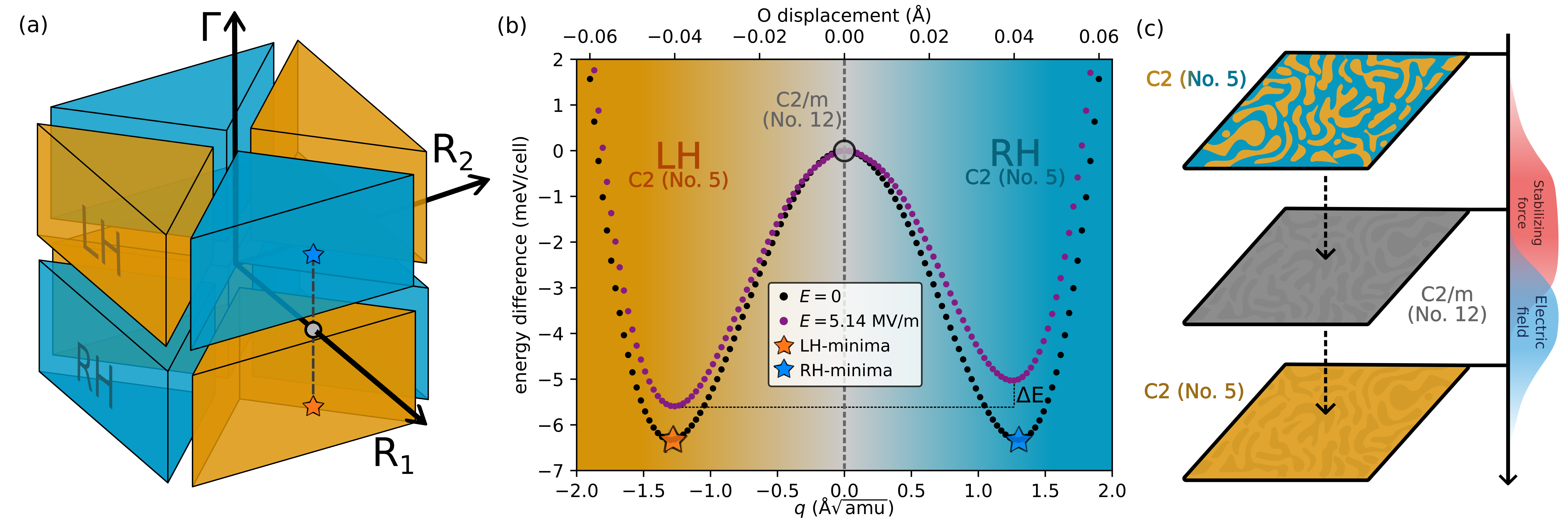}
    \caption{
        \textbf{Chiral selectivity mechanism in $\mathbf{NbOCl_2}$.}
        \textbf{a.} Representation of the three-dimensional order parameter space $(R_{1},R_2,\Gamma)$ containing all the space-groups described in Fig. \ref{fig:1}(g).
        The orange/blue colors are used to represent the left/right handed  enantiomers.
        Both enantiomers are always divided by an achiral phase (that we represented as the empty space).
        The high-symmetry phase is located at the center of the cube, the intermediate phase is located across the $R_1$ and $R_2$ axes (represented by a gray dot), and both enantiomers of the low symmetry chiral phase are represented as the orange/blue stars.
        \textbf{b.} The energies of the order parameter path (dotted line in \textbf{a}) mediating both enantiomers in the absence and the presence of an electric field.
        In the absence of an electric field, both enantiomers have the same energy, however, the electric field lifts the degeneracy and energetically favours one of them.
        \textbf{c.} Sketch of the chirality reversal process, where temperature lowers the energy barrier between enantiomers (increasing the stability of the intermediate $C2/m$ phase), while an electric field biases the potential toward the desired handedness.
    }
    \label{fig:4}
\end{figure*}

Thus far, we have established a connection between both the achiral \(Immm\) phase and the chiral \(C2\) phase, each experimentally reported, through an intermediate achiral \(C2/m\) phase, which is prone to stabilization.
The next step is to investigate whether this intermediate $C 2/m$ phase can bridge the two enantiomers and, ultimately, whether it can favor one over the other.
To address this, we explore the space groups resulting from the three-dimensional order parameter space spanned by $(R_1, R_2, \Gamma)$.
As shown in Fig. \ref{fig:1}(g), there are seven possible space groups ($C 2/m$, $C m m m$, $P 2/m$, $I m m 2$, $C2$, $C m m 2$, $P 2$), two of which are chiral.
By applying symmetry operations from the $I m m m$ phase to the order parameter space, we can identify sections related by proper operations (which connect rotated versions of the same crystal) as well as sections related by improper operations (which connect the two enantiomers) (see supplementary information for details \cite{supplementary}).
Using this information, we divide the order parameter space, assigning a handedness to each chiral section as depicted in Fig. \ref{fig:4}(a).
As expected, domains with different chiralities are always separated by an achiral phase (thus the gaps in Fig. \ref{fig:4}(a)).
Interestingly, the $C 2$ phase is located near a chirality interface that can be crossed by condensing the $\Gamma_{1}^{-}$ mode in the opposite direction.

As previously discussed, the energy barrier between the enantiomers is only a few meV (Fig. \ref{fig:2}(b)), indicating that a transition between them is feasible.
In the $C2/m$ structure, mirror and inversion symmetries relate $(R_1, R_2, \Gamma)$ to $(R_1, R_2, -\Gamma)$, so breaking these symmetries will lift the degeneracy and energetically favour one enantiomer.
To test this, we recomputed the BOES of the $\Gamma_1^{-}$ mode under an electric field $\mathbf{E}$ of $5.14\ \mathrm{MV/m}$ applied along the $a^{*}$ direction, which is nearly perpendicular to the mirror plane.
As shown in Fig.~\ref{fig:4}(b), the field breaks mirror symmetry and favours the left-handed enantiomer, while reversing $\mathbf{E}$ would favour the right-handed one.
Combined with the small energy barrier, this provides a viable mechanism to select a desired handedness by manipulating an external field, as illustrated in Fig.~\ref{fig:4}(c).
Assuming the system initially hosts an equal population of both enantiomers, applying the electric field breaks the degeneracy between them.
Utilizing temperature or pressure as stabilizing factors effectively lowers the energy barrier by exploiting anharmonic effects, thereby facilitating the transition from one enantiomer to the other.
Subsequently, by reducing these stabilizing forces and removing the electric field, the population of the desired enantiomer becomes highly biased.

In summary, we have investigated the emergence of chirality in the layered $\mathrm{NbOX_2}$ family of crystals, tracing the transition from the achiral $I m m m$ phase to the chiral $C 2$ phase via an intermediate achiral $C 2/m$ phase for $\mathrm{NbOCl_2}$.
Through the analysis of the Born-Oppenheimer energy surface, we identified the $C 2/m$ phase as a low-energy, unstable achiral phase, which can be stabilized through external pressure or anharmonic effects.
We demonstrated that the chiral $C 2$ phase exhibits remarkable features, including flat Niobium $d$-orbital bands near the Fermi level, corresponding to an obstructed atomic limit, and hosting non-trivial surface states under specific cleavage conditions.
Finally, we proposed a mechanism for selectively controlling the handedness of the chiral $C 2$ phase using an external electric field, exploiting the small energy barrier between enantiomers and the effects of either pressure or thermal fluctuations.
This work not only provides fundamental insights into the mechanisms driving chirality in this family of materials but also establishes a framework for tailoring enantiomeric populations through external stimuli.

\begin{acknowledgments}
    M.G.-A. was supported by Jane and Aatos Erkko Foundation, Keele Foundation, and Magnus Ehrnrooth Foundation as part of the SuperC collaboration.
    I.E. and M.G.-A acknowledge the Spanish Ministerio de Ciencia e Innovaci\'on grants PID2022-142861NA-I00. I.E. acknowledges the Department of Education, Universities and Research of the Eusko Jaurlaritza and the University of the Basque Country UPV/EHU (Grant No. IT1527-22).
    M.G.V. thanks support to PID2022-142008NB-I00 project funded by  MICIU/AEI/10.13039/501100011033 and FEDER, UE, Canada Excellence Research Chairs Program for Topological Quantum Matter, NSERC Quantum Alliance France-Canada, and to Diputación Foral de Gipuzkoa Programa Mujeres y Ciencia.
    We also acknowledge funding from the EU NextGenerationEU/PRTR-C17.I1, as well as by the IKUR Strategy under the collaboration agreement between Ikerbasque Foundation and DIPC on behalf of the Department of Education of the Basque Government. This work has also been  funded by the Ministry of Economic Affairs and Digital Transformation of the Spanish Government through the QUANTUM ENIA project call – Quantum Spain project, and by the European Union through the Recovery, Transformation and Resilience Plan – NextGenerationEU within the framework of the Digital Spain 2026 Agenda.
\end{acknowledgments}

\clearpage %
\bibliographystyle{apsrev4-1} %
\bibliography{bibtex.bib}
\end{document}


\date{May 14, 2025}
\title{
    \textit{Supplementary Material for }\\
    Emergent chirality and enantiomeric selectivity in layered NbOX$_2$ crystals
}

\author{Martin Gutierrez-Amigo}
\thanks{Corresponding author - Martin Gutierrez-Amigo \\
    martin.gutierrezamigo@aalto.fi
}

\affiliation{Department of Applied Physics, Aalto University School of Science, FI-00076 Aalto, Finland}

\author{Claudia Felser}
\affiliation{Max Planck Institute for Chemical Physics of Solids, 01187 Dresden, Germany}

\author{Ion Errea}
\affiliation{Centro de Física de Materiales (CFM-MPC), CSIC-UPV/EHU, 20018 Donostia/San Sebasti\'an, Spain}
\affiliation{Fisika Aplikatua Saila, Gipuzkoako Ingeniaritza Eskola, University of the Basque Country (UPV/EHU), 20018 Donostia/San Sebasti\'an, Spain}
\affiliation{Donostia International Physics Center (DIPC), 20018 Donostia/San Sebasti\'an, Spain}

\author{Maia G. Vergniory}
\affiliation{Département de Physique et Institut Quantique, Université de Sherbrooke, Sherbrooke, J1K 2R1 Québec, Canada.}
\affiliation{Donostia International Physics Center (DIPC), 20018 Donostia/San Sebasti\'an, Spain}
\affiliation{Regroupement Qu\'eb\'ecois sur les Mat\'eriaux de Pointe (RQMP), Quebec H3T 3J7, Canada}

\maketitle
\newpage

\section{Details of first-principle calculations}

First-principles density functional theory (DFT) calculations were performed using the \textsc{Quantum Espresso} package \cite{giannozzi_2009,giannozzi_2017}.
We employed the generalized gradient approximation (GGA) with the Perdew-Burke-Ernzerhof (PBE) exchange-correlation functional \cite{perdew_1996}, in combination with projector-augmented wave (PAW) pseudopotentials \cite{kresse_1999} generated by Dal Corso \cite{corso_2014}.
We considered $13 / 6$ valence electrons for niobium and oxygen, and  $7 / 17 / 17$ chlorine, bromide and iodine respectively.
Plane-wave energy cutoffs of 100 Ry for the wavefunctions and 1000 Ry for the density were chosen, along with a Methfessel-Paxton smearing \cite{methfessel_1989} of 0.02 Ry.
A $12 \times 12 \times 6$ $k$-point grid was used for the $Im m m$ phase, while a $12 \times 12 \times 6$ grid was used for $C2$ and $C 2/m$ phases due to cell doubling.

Due to the layered nature of these compounds (see Fig. 1(a)), we compared different implementations of Van der Waals (VdW) corrections during structural relaxations.
The relaxations, which did not include spin-orbit coupling (SOC), were performed starting from the experimental lattice parameters reported in \cite{hillebrecht_1997}.
Relaxations were stopped when the forces on atoms were below 0.001 Ry/au, and the final pressure was below 0.02 kbar.

\begin{figure}[h]
    \centering
    \includegraphics[width=\linewidth]{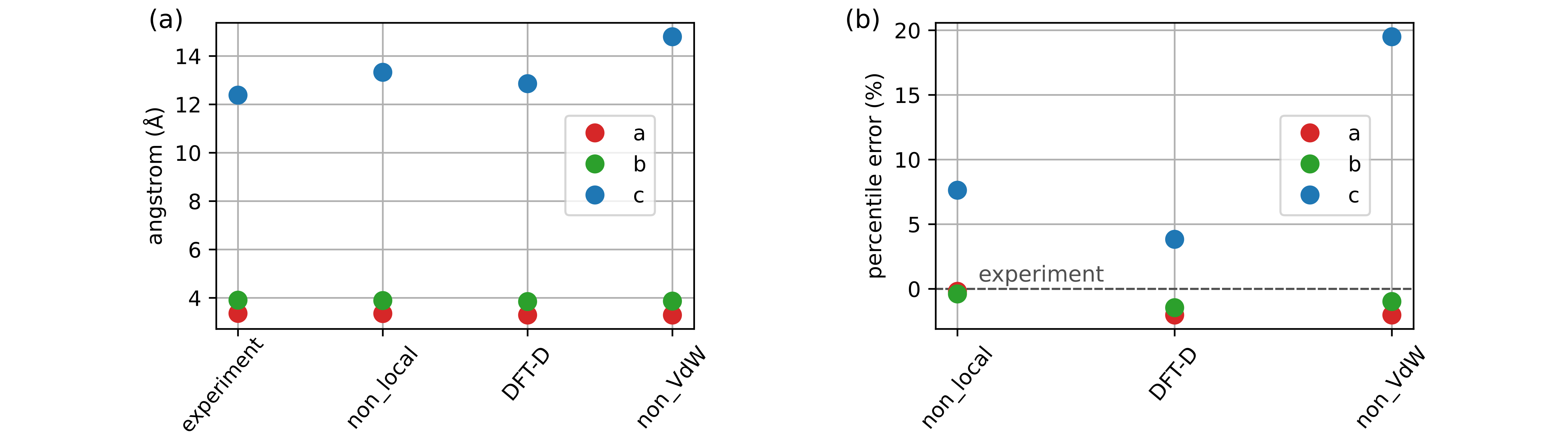}
    \caption{
        \textbf{Comparison of Van der Waals implementations for $\mathrm{NbOCl_2}$.}
        \textbf{a.} Experimental lattice parameters \cite{hillebrecht_1997} and relaxed lattice parameters for the $Immm$ phase using a non-local functional, the Grimme semi-empirical approach, or by neglecting Van der Waals interactions.
        \textbf{b.} Percentile error respect to the experimental data.
    }
    \label{fig:sup_lat}
\end{figure}

We tested three approaches for incorporating VdW interactions: the non-local functional developed by Dion \textit{et al.} \cite{dion_2004}, Grimme's semi-empirical correction (DFT-D2) \cite{grimme_2006}, and a baseline case without any VdW corrections.
The resulting relaxed lattice parameters were compared against the experimental values \cite{hillebrecht_1997}, as shown in Fig.~\ref{fig:sup_lat}.
Grimme's approach provided the best agreement, particularly for the lattice vector parallel to the Van der Waals direction, suggesting it most accurately reproduces the VdW interactions.
Therefore, we adopted the Grimme correction for all subsequent calculations.

To compute the electronic band structures and Fermi surfaces, SOC was included in the calculations.
A finer, non-self-consistent $20 \times 20 \times 20$ $k$-point grid was used in the Fermi surface and electron localization function calculations.

Harmonic phonon spectra were obtained using density functional perturbation theory (DFPT) \cite{baroni_2001}.
Phonon calculations were performed on a $4 \times 4 \times 4$ grid for the $Immm$ phase and on a $2 \times 2 \times 2$ grid for the $C2/m$ and $C2$ phases.
SOC was not included in the phonon calculations. The spectra were obtained by Fourier interpolation.

\subsection*{Born-Oppenheimer energy surface calculations}
To compute the Born-Oppenheimer (BO) energy surfaces shown in Figs. 1, 2, and 4 of the main manuscript, we calculate the total energy by displacing the ions according to the corresponding active phonon modes.
In the harmonic approximation, the displacement $\mathbf{u}$ of an atom $s$ in the unit cell at position $\mathbf{R}$ is given by:
\begin{equation} \label{eq:disp}
    {u}^{\alpha}_s(\mathbf{R})=\mathrm{Re}\{ \sum_{\mu \mathbf{k}} q_\mu (\mathbf{k}) \frac{\mathbf{\varepsilon}_{\mu s}^{\alpha}(\mathbf{k})}{\sqrt{M_s}}e^{i\mathbf{k}\cdot\mathbf{R}} \}
\end{equation}

Here, $\alpha$ denotes a Cartesian coordinate, $\mu$ labels the phonon mode, $M_s$ is the mass of atom $s$, $\varepsilon_{\mu s}^{\alpha} (\mathbf{k})$ is the polarization vector, and $q_\mu(\mathbf{k})$ is the order parameter associated with the $\mu$ mode at wave vector $\mathbf{k}$.
By plotting the energy as a function of the order parameter $q_{\mu}(\mathbf{k})$, we obtain the BO energy surface along the specified direction in the order parameter space.
For Fig. 1(e), where the energy surface is plotted against two distinct order parameters corresponding to the $R_1$ and $R_2$ wavevectors, we sampled the order parameter space on a regular $21 \times 21$ grid and applied cubic interpolation to refine it to a denser $201 \times 201$ grid.

\subsection*{Surface state ARPES simulations}
To generate the surface states calculations shown in Figs. 3 and \ref{fig:sup_surf}, we employed the Wannierization procedure implemented in Wannier90 \cite{mostofi_2014}, along with WannierTools \cite{wu_2018}.
First, we obtained a tight-binding model with a Wannierization considering chlorine and oxygen $p$-orbitals as well as niobium $d$-orbitals, all of them centered in their corresponding atomic sites.
Then, we calculated the surface spectrum by a method of iterative Green's function as implemented in WannierTools \cite{wu_2018} for a slab of 10 unit cells.

\subsection*{Electron localization function}
The electron localization function (ELF) is a useful tool for analyzing electron localization in real space \cite{becke_1990,savin_1997} and was employed in this study to investigate the layered structure and identify potential natural cleavage surfaces.
To define the ELF, we begin by introducing the ratio $\chi$, given as:
\begin{equation} \label{eq:ELF1} \chi = \frac{t_{p}}{t_{\mathrm{HEG}}}, \end{equation}
where $t_{p}$ represents the Pauli kinetic energy density, and $t_{\mathrm{HEG}}$ corresponds to the kinetic energy density of a homogeneous electron gas \cite{kohout_1996}.
The ELF is then defined as:
\begin{equation} \label{eq:ELF2} \mathrm{ELF} = \frac{1}{1 + \chi^{2}}, \end{equation}
ensuring that its values range between 0 and 1.
Regions where the ELF approaches 1 indicate highly localized electrons with low Pauli kinetic energy, often corresponding to strong chemical bonds, whereas an ELF value of 0.5 indicates a distribution similar to that of a homogeneous electron gas with the same local electron density.
Consequently, areas with low ELF can be associated with natural cleavage surfaces.

\section{Multimode order-parameter analysis}
The phonon spectrum of the high-symmetry $Immm$ phase (Fig. 1(d)) exhibits multiple instabilities, obscuring the identification of the modes responsible for the CDW.
These instabilities imply a multidimensional order parameter, and determining the energy-minimizing direction in this high-dimensional space is nontrivial.
However, the problem is significantly constrained by symmetry considerations.
The CDW transition leads to a low-symmetry $C2$ structure (No.~5) \cite{seifert_1981,guo_2023}, which contains eight atoms per primitive cell, twice as many as in the $Immm$ phase, which restricts the set of symmetry-allowed mode combinations.
The negative phonon frequencies correspond to negative second derivatives of the Born-Oppenheimer energy surface (BOES) with respect to displacement fields, indicating potential pathways to minimize the energy.
We explore combinations of the unstable phonon modes at $R$, $\Gamma$, $X$, $T$, and $W$ points, where $\left[ R, T, W \right]$ have two symmetry-equivalent directions (a star of two), and $\left[ X, \Gamma \right]$ have one (a star of one).
These modes transform under the irreducible representations ${R_{2}^{-}, \Gamma_{3}^{-}, X_{3}^{-}, T_{2}^{-}, W_{4}}$, resulting in a total of eight unstable modes and an 8-dimensional order parameter.
While an exhaustive exploration of this space is computationally prohibitive, symmetry and cell-size constraints help narrow the possibilities.
The condensation of $W_4$ necessarily yields structures with at least 16 atoms per primitive cell, making it incompatible with the observed $Immm \rightarrow C2$ transition.
Among the remaining modes, all modes except $\Gamma_{3}^{-}$ double the unit cell, preventing their simultaneous condensation.
This leaves three possible combinations of $\Gamma_{3}^{-}$ with either $R_{2}^{-}$, $X_{3}^{-}$, or $T_{2}^{-}$ modes.
Only the combination of $R_{2}^{-}$ with $\Gamma_{3}^{-}$ produces the $C2$ structure (see Fig. 1(g)), identifying it as the likely pathway connecting the high- and low-symmetry phases.
This result is further supported by Fermi surface nesting analysis: the $R$ vector precisely connects two Fermi sheets in Fig. \ref{fig:sup_fermi}(a), producing a strong nesting function and likely enhancing electron–phonon coupling (details in section \ref{sec:fermi}).
We thus identify a consistent, symmetry-allowed pathway between the $Immm$ and $C2$ phases involving only the $\Gamma$ and $R$ modes, compatible with the observed atomic multiplicities and in agreement with prior reports \cite{hillebrecht_1997,seifert_1981,jia_2019,guo_2023,fu_2024,zhang_2024}.

\section{Fermi surface nesting}\label{sec:fermi}
A useful indicator of lattice instabilities is the presence of wavevectors that nest large portions of the Fermi surface.
This is because the small phonon energies and momentum conservation require that electrons near the Fermi level be connected by a phonon with the appropriate wavevector, ensuring momentum conservation and enabling meaningful electron-phonon coupling.
While nesting alone does not guarantee lattice instabilities and significant electron-phonon coupling matrix elements between these electrons and phonons is also essential, it is nevertheless a necessary condition for meaningful electron-phonon coupling.

\begin{figure}[H]
    \centering
    \includegraphics[width=\linewidth]{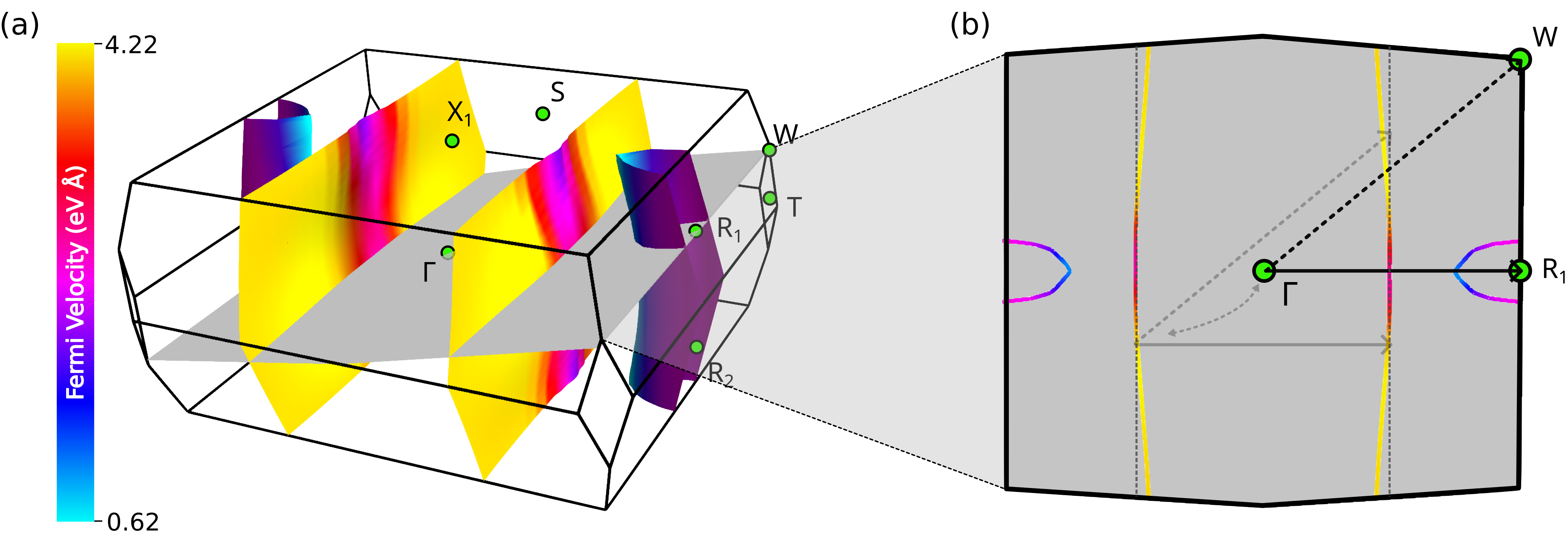}
    \caption{
        \textbf{Fermi surface and nesting for the $\mathrm{NbOCl_2}$ in the $Im m m$ phase.}
        \textbf{a.} Full Fermi surface with the $\Gamma$ and $R_1$ high-symmetry points highlighted.
        The gray section indicates a two-dimensional plane containing both high-symmetry points (HSPs).
        \textbf{b.} Energy cut of the Fermi surface across the section shown in (a).
        The $R_1$ vector is highlighted in black, along with its displaced counterparts (in red), illustrating how the $R_1$ wavevector nests the two large Fermi sheets.
    }
    \label{fig:sup_fermi}
\end{figure}

In this case, as shown in Fig.~\ref{fig:sup_fermi}(a), the Fermi surface exhibits strong two-dimensional character, with the non-dispersive direction aligned parallel to the Van der Waals stacking direction.
Furthermore, the presence of two large, flat Fermi sheets, reminiscent of a one-dimensional crystal, makes them particularly prone to high nesting with the appropriate $\mathbf{q}$-wavevector phonon due to their flatness.
In this case, the $R_1$ and $R_2$ wavevectors perfectly nest the Fermi sheets, as illustrated in Fig. \ref{fig:sup_fermi}(b).
Additionally, the $T$ and $W$ wavevectors also exhibit strong nesting of the same two Fermi sheets, which correlates with the pronounced instabilities observed in the phonon spectra shown in Fig. 1(d).

\section{Topological quantum chemistry analysis}
\begin{figure}[h]
    \centering
    \includegraphics[width=0.7\linewidth]{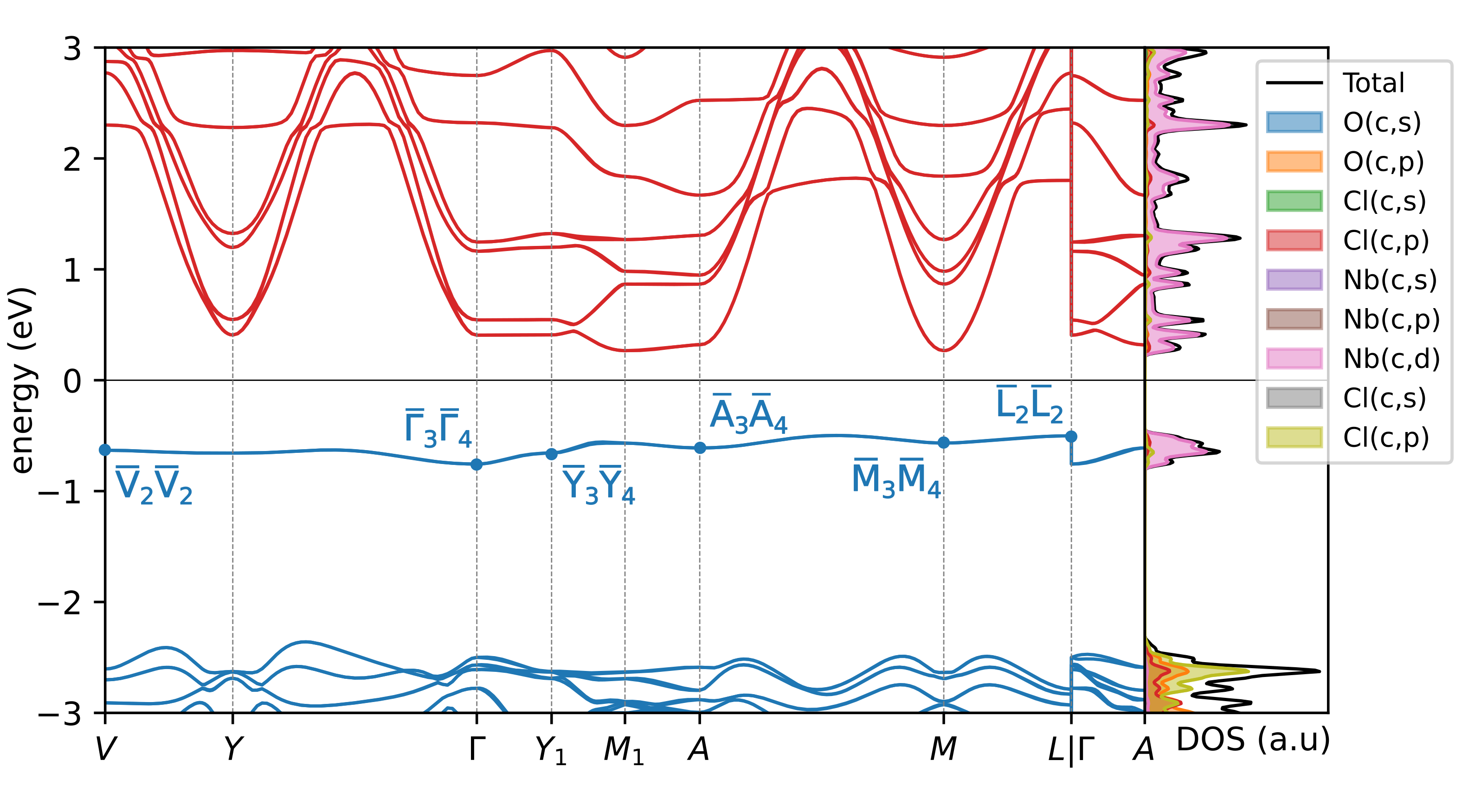}
    \caption{
        \textbf{Band structure for $\mathbf{NbOCl_2}$ in the $\mathbf{C 2}$ phase.}
        The irreducible representations at the high-symmetry points are shown in blue for the band corresponding to an obstructed atomic limit.
    }
    \label{fig:sup_bands_C2}
\end{figure}

Besides the interest of the $C 2$ phase due to it's chiral selectivity, its electronic spectra is also non-trivial.
According to the symmetry content of the band structure, the $C2$ phase (see Fig. \ref{fig:sup_bands_C2}) is classified as an obstructed atomic limit (OAL).
The irreducible representations, and thus the symmetry properties of the Bloch functions across the last pair of valence bands, which lie about 0.7 eV below the Fermi level (see Fig. \ref{fig:sup_bands_C2}), correspond to those of the elementary band representation \cite{cano_2018,bradlyn_2017} that would be induced from a pair of Wannier functions centered at the 2a Wyckoff position transforming under the site-symmetry group representation $\bar{E}^{1}\bar{E}^{2}$.

\begin{figure}[H]
    \centering
    \includegraphics[width=0.7\linewidth]{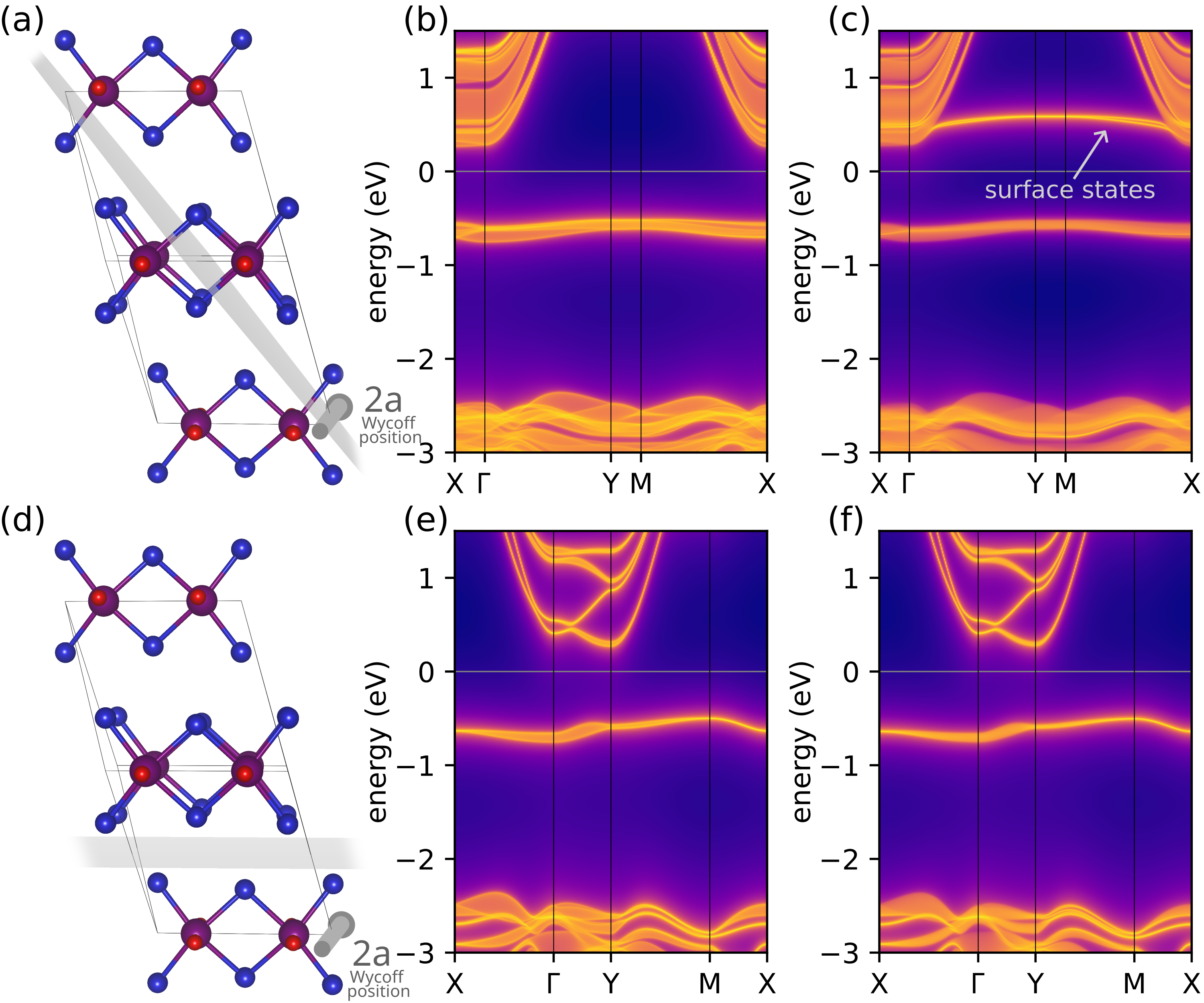}
    \caption{
        \textbf{Surface states for $\mathbf{NbOCl_2}$ in the $\mathbf{C 2}$ ground state.}
        \textbf{a,d.} Primitive cell of $\mathrm{NbOCl_2}$ in space group $C 2$ with the 112 (a) and 110 (d) Miller planes highlighted, along with the 2a Wyckoff position, which corresponds to any point along the axis indicated by the gray arrow.
        \textbf{b,e.} Bulk density of states along a reciprocal space path constrained to the sections shown in (a) and (d), respectively.
        \textbf{c,f.} Surface density of states for the 112 and 110 surfaces.
        In panel (c), surface states absent in the bulk (b) can be clearly observed.
    }
    \label{fig:sup_surf}
\end{figure}

This suggests that in real space, the Wannier functions representing the electrons are localized at Wyckoff positions not occupied by the ions, as shown in Fig. \ref{fig:sup_surf}(a,d).
When a surface cuts through these Wyckoff positions, it will exhibit symmetry protected surface states \cite{cano_2022,ma_2023,xu_2024a}, as shown in Fig. \ref{fig:sup_surf}(c).
However, our previous analysis of the ELF suggests that the natural cleavage surface is likely the $(110)$ plane, which does not intersect the 2a Wyckoff position.
Consequently, this surface lacks the expected surface states, as it does not cut through any of the Wannier centers (see Fig. \ref{fig:sup_surf}(d)).

\section{Transformation of the $\mathbf{(R_1,R_2,\Gamma)}$ order parameter under symmetry operations.}
The $(R_1,R_2,\Gamma)$ order parameter represents the condensation of three independent phonons that transform under irreducible representations ($R_{2}^{-},R_{2}^{-},\Gamma _{3}^{-})$ of the high-symmetry space group $I m m m$.
The symmorphic representations of the $I m m m$ group in the $(R_1,R_2,\Gamma)$ basis can be obtained from the generators:

The $(R_1, R_2, \Gamma)$ order parameter represents the condensation of three independent phonons that transform under the irreducible representations $(R_{2}^{-}, R_{2}^{-}, \Gamma _{3}^{-})$ of the high-symmetry space group $I m m m$.
The representations of the symmorphic operations of $I m m m$ group in the $(R_1, R_2, \Gamma)$ basis can be obtained from the generators:

\begin{equation} \label{eq:OP_sym}
    1=\begin{pmatrix}
        1 & 0 & 0\\
        0 & 1 & 0\\
        0 & 0 & 1
    \end{pmatrix}
    ;
    I=\begin{pmatrix}
        -1 & 0 & 0\\
        0 & -1 & 0\\
        0 & 0 & -1
    \end{pmatrix}
    ;
    2_{001}=\begin{pmatrix}
        0 & -1 & 0\\
        -1 & 0 & 0\\
        0 & 0 & -1
    \end{pmatrix}
    ;
    2_{010}=\begin{pmatrix}
        -1 & 0 & 0\\
        0 & -1 & 0\\
        0 & 0 & -1
    \end{pmatrix}.
\end{equation}
The structure resulting from the condensation of phonons corresponding to a given order parameter will retain the symmetries under which that order parameter remains invariant.
Moreover, all order parameters related by the original symmetries correspond to crystals with the same energy, though not necessarily the same crystal structure.
In particular, when an order parameter is not invariant under any improper symmetry operations, the resulting crystal must be chiral, with its enantiomer related to another order parameter connected via an improper symmetry operation.

It can be shown that the only $(R_1, R_2, \Gamma)$ order parameters that are not invariant under any improper symmetry operations take one of the following forms: $(a,0,c)$, $(a,a,c)$, or $(a,b,c)$.
These correspond to the space groups $C_2$ (No. 5), $C m m 2$ (No. 35), and $P_2$ (No. 3), respectively.

Thus, we can divide the order parameter space based on handedness as follows:
First, we assign a handedness to a particular chiral order parameter of the form $(a,b,c)$.
Then, by successively applying inversion $I$ and the $C_{2z}$ rotation (given in Eq. \ref{eq:OP_sym}), we can generate all related order parameters and determine their handedness.
Since inversion flips the handedness, while $C_{2z}$ merely rotates the structure without changing handedness, we can systematically classify the entire order parameter space.
As a result, the division of order parameter space into distinct left and right-handed domains is shown in Fig. 4(a).

\section{Systematic analysis for all compounds in $\mathbf{NbOX_2}$ family}

So far, all analyses and calculations have been performed on $\mathrm{NbOCl_2}$ as a representative of the $\mathrm{NbOX_2}$ family, where $X$ can be chlorine, bromine, or iodine.
However, an identical systematic analysis has also been carried out for $\mathrm{NbOBr_2}$ and $\mathrm{NbOI_2}$, yielding equivalent qualitative results that support the same reasoning developed above, with only minor quantitative variations.

\begin{figure}[H]
    \centering
    \includegraphics[width=\linewidth]{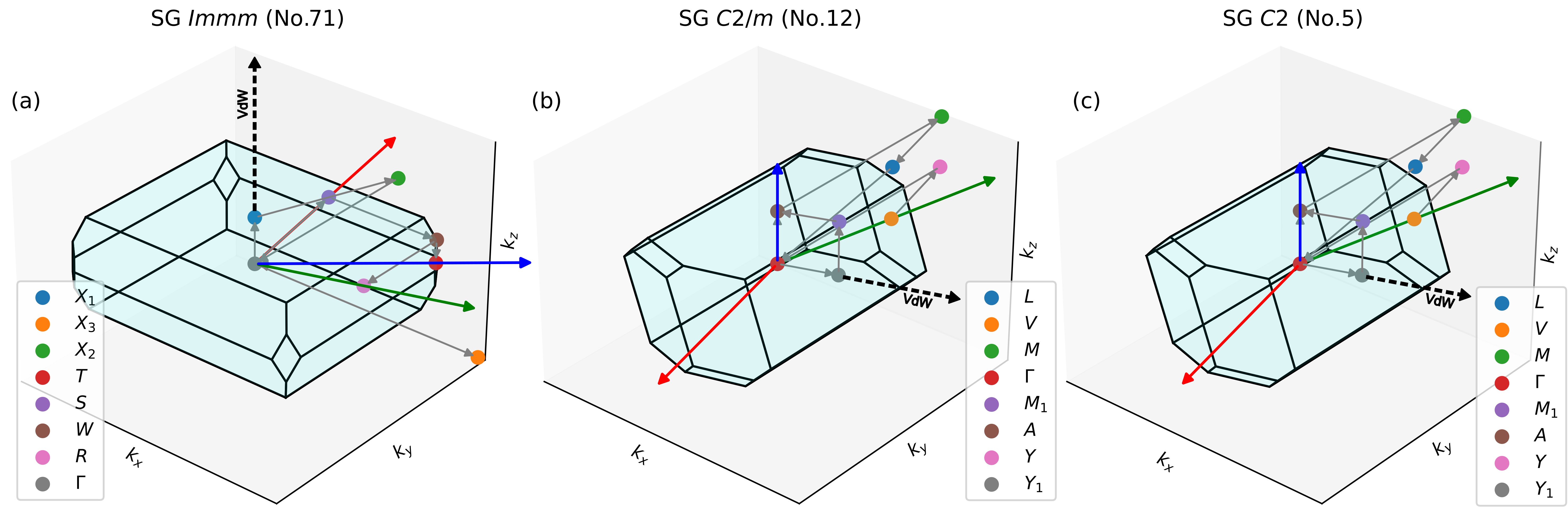}
    \caption{
        \textbf{Brillouin zones for all space groups.}
        Brillouin zones and high-symmetry points are presented for all space groups.
        The reciprocal base vectors \(a^{*}\), \(b^{*}\), and \(c^{*}\) are colored red, green, and blue, respectively.
        The high-symmetry paths used for phonon and band spectra are illustrated with gray arrows, and the van der Waals direction is highlighted with a dashed black line.
    }
    \label{fig:sup_BZs}
\end{figure}

In this section, we compile all calculations and results for $\mathrm{NbOCl_2}$, $\mathrm{NbOBr_2}$, and $\mathrm{NbOI_2}$ across the three phases $Immm$, $C2/m$, and $C2$.
For each phase, the corresponding Brillouin zones, together with the high-symmetry paths used, are shown in Fig.~\ref{fig:sup_BZs}; the relevant van der Waals direction is also highlighted in each case.

In Subsection~\ref{sec:structures}, all Wyckoff positions defining the crystal structures for the three compounds in the three phases are provided in Tables~\ref{tab:Cl}, \ref{tab:Br}, and \ref{tab:I}.

\begin{figure}[H]
    \centering
    \includegraphics[width=\linewidth]{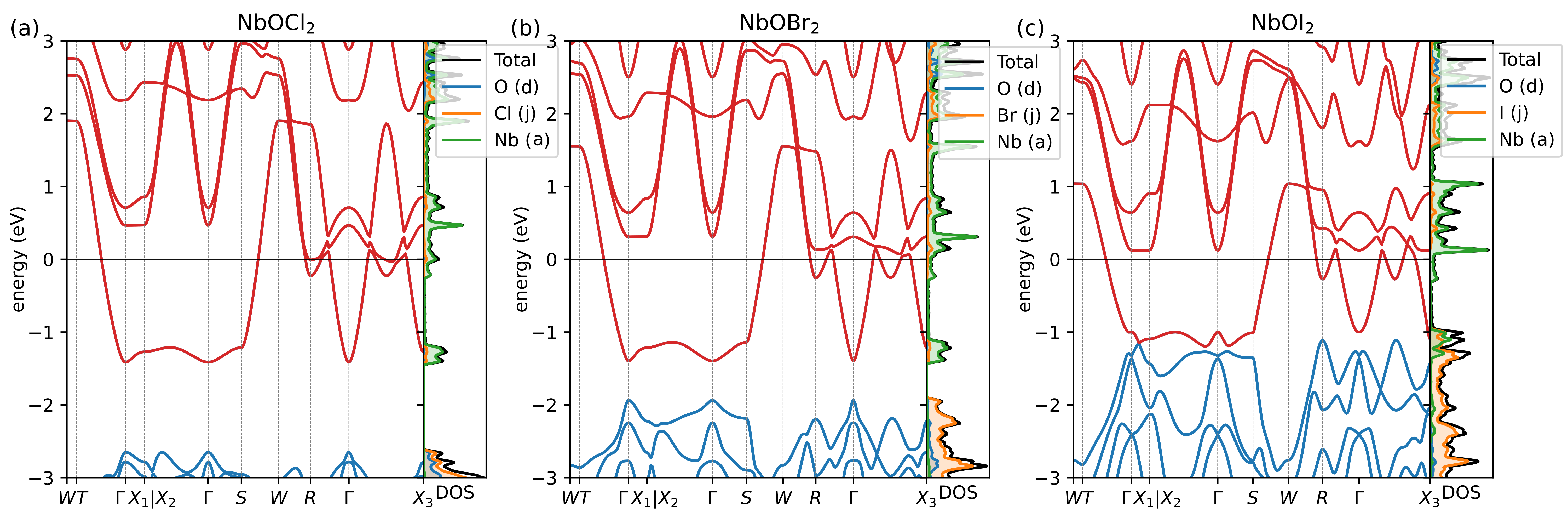}
    \caption{
        \textbf{Bandstructures for SG $\mathbf{Im m m}$ (No.71).}
        Band structures and projected density of states for $\mathrm{NbOCl_2}$, $\mathrm{NbOBr_2}$, and $\mathrm{NbOI_2}$ are shown in panels $(a)$, $(b)$, and $(c)$, respectively.
        Labels indicate the Wyckoff position of each contribution.
    }
    \label{fig:sup_SG71_bands}
\end{figure}

\begin{figure}[H]
    \centering
    \includegraphics[width=\linewidth]{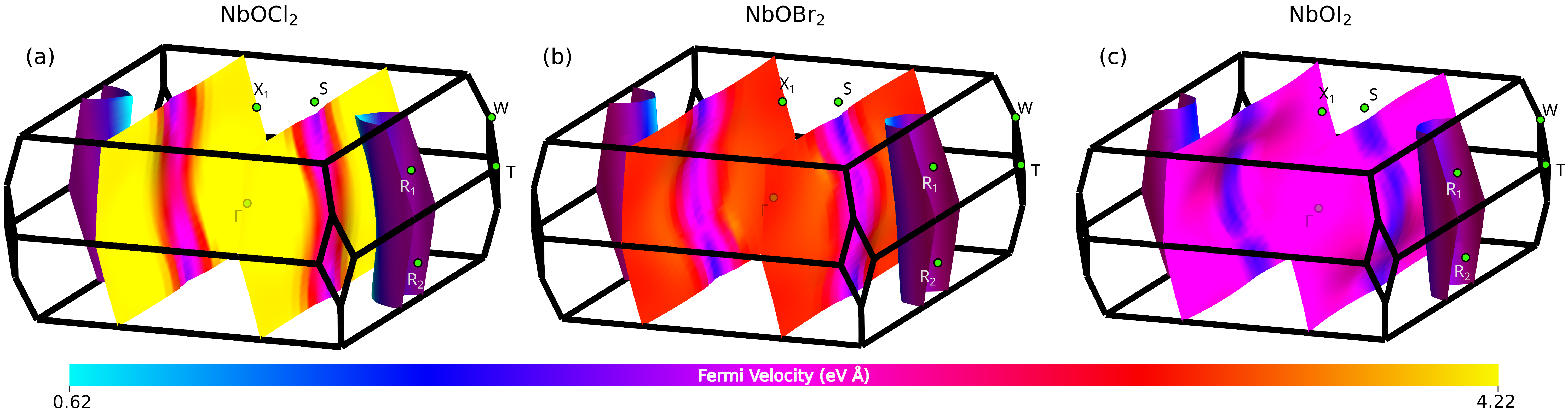}
    \caption{
        \textbf{Fermi surfaces for SG $\mathbf{Im m m}$ (No.71).}
        Fermi surfaces for $\mathrm{NbOCl_2}$, $\mathrm{NbOBr_2}$, and $\mathrm{NbOI_2}$ are shown in panels $(a)$, $(b)$, and $(c)$, respectively.
        In all cases, the same nesting arguments discussed in Section \ref{sec:fermi} are applicable.
    }
    \label{fig:sup_SG71_fermi}
\end{figure}

\begin{figure}[H]
    \centering
    \includegraphics[width=\linewidth]{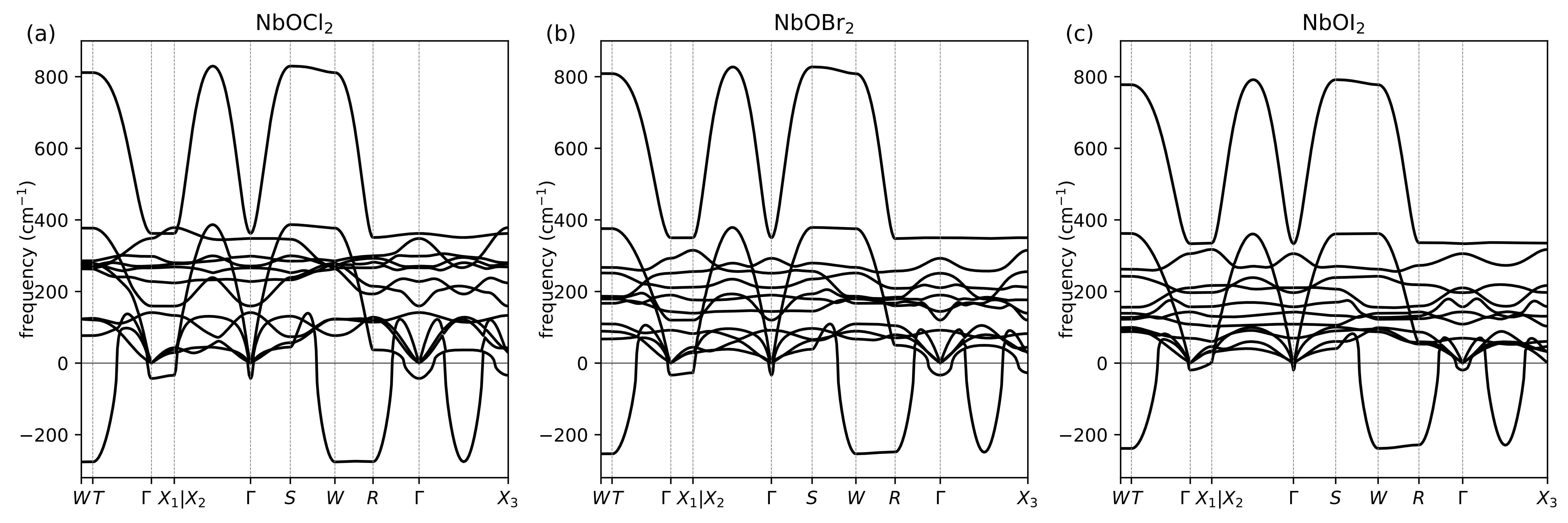}
    \caption{
        \textbf{Phonons for SG $\mathbf{Im m m}$ (No.71).}
        Phonon spectra for $\mathrm{NbOCl_2}$, $\mathrm{NbOBr_2}$, and $\mathrm{NbOI_2}$ are shown in panels $(a)$, $(b)$, and $(c)$, respectively.
        The same instabilities are present in all compounds.
    }
    \label{fig:sup_SG71_phonons}
\end{figure}

\begin{figure}[H]
    \centering
    \includegraphics[width=\linewidth]{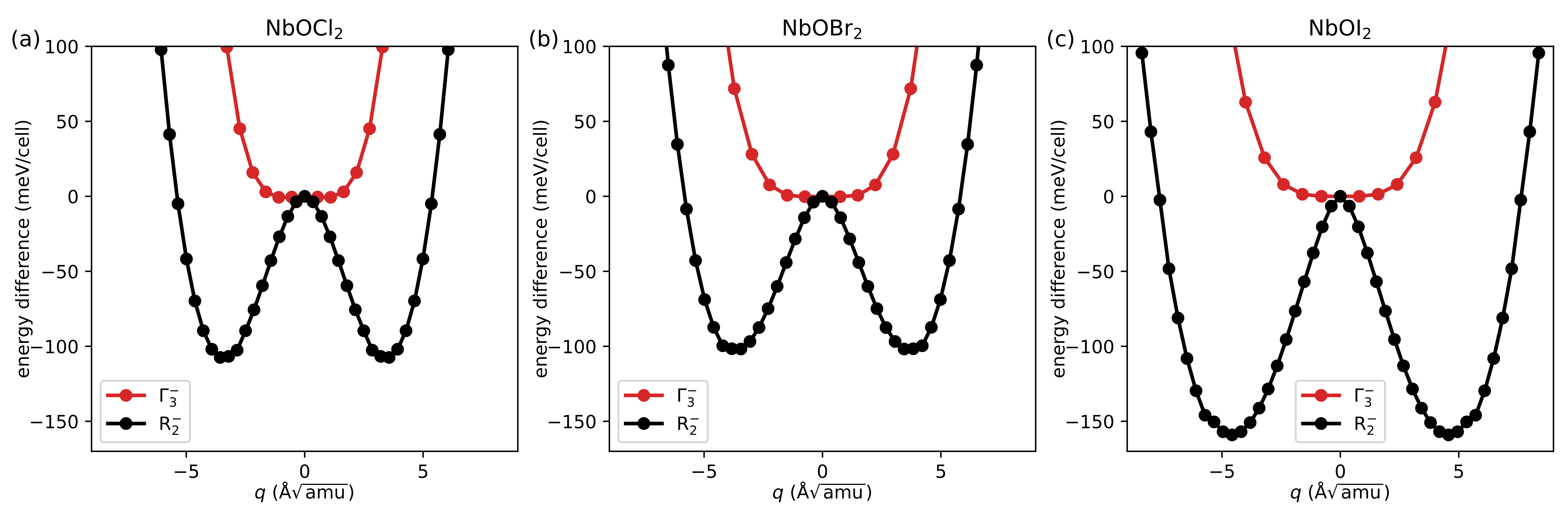}
    \caption{
        \textbf{$\mathbf{\Gamma_{3}^{-}}$ and $\mathbf{R_{2}^{-}}$ BOES for SG $\mathbf{Immm}$ (No.71).}
        Computed Born-Oppenheimer energy surfaces as the high-symmetry structure is distorted either parallel to the $\Gamma_{3}^{-}$ or $(R_{2}^{-},R_{1}^{-})$ modes for $\mathrm{NbOCl_2}$, $\mathrm{NbOBr_2}$, and $\mathrm{NbOI_2}$ are shown in panels $(a)$, $(b)$, and $(c)$, respectively.
        In all compounds, the $R_2^{-}$ instability dominates over the $\Gamma_3^{-}$ instability.
    }
    \label{fig:sup_SG71_BOES1}
\end{figure}

\begin{figure}[H]
    \centering
    \includegraphics[width=\linewidth]{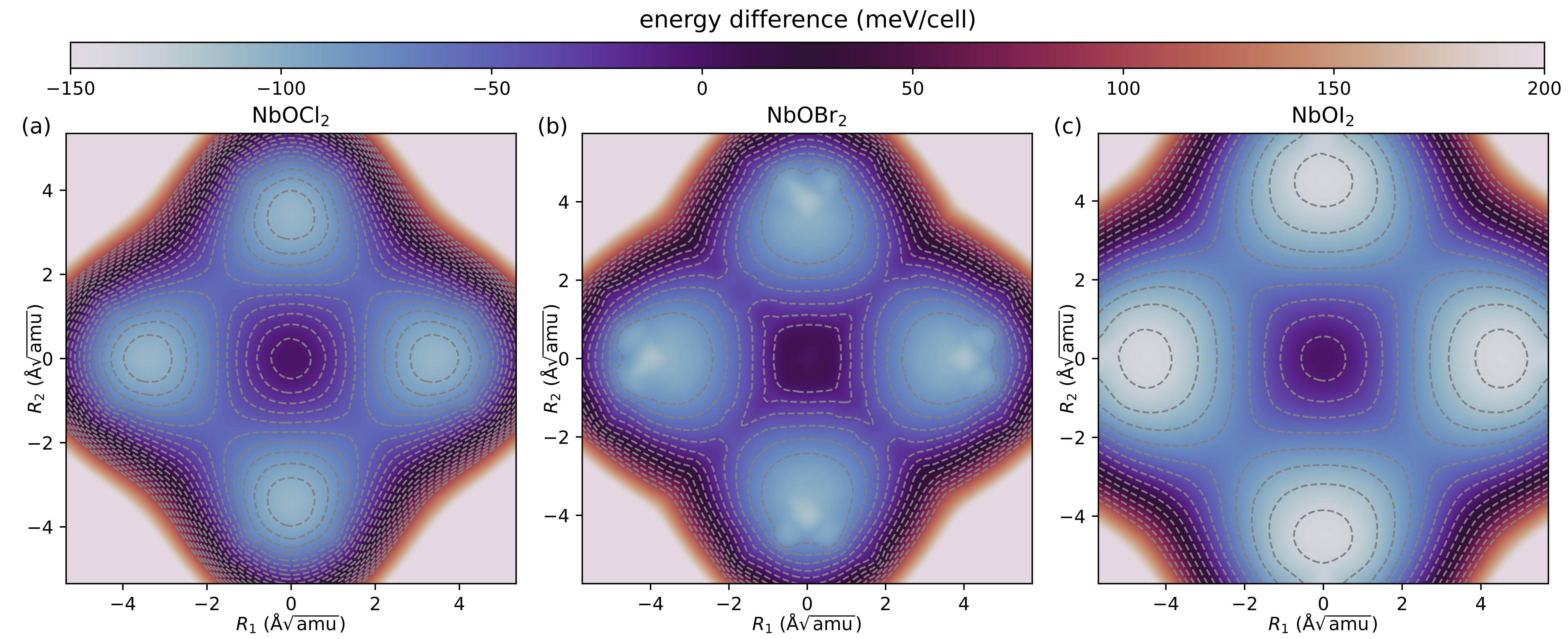}
    \caption{
        \textbf{$\mathbf{\left\{ R_1,R_2 \right\}}$ BOES for SG $\mathbf{Immm}$ (No.71).}
        Computed 2D Born-Oppenheimer energy surface as the high-symmetry structure is distorted by $\left\{ R_1,R_2 \right\}$ modes for $\mathrm{NbOCl_2}$, $\mathrm{NbOBr_2}$, and $\mathrm{NbOI_2}$ are shown in panels $(a)$, $(b)$, and $(c)$, respectively.
        In all compounds, the global minima correspond to the isolated condensation of either $R_1$ or $R_2$ modes.
    }
    \label{fig:sup_SG71_BOES2}
\end{figure}

Subsection~\ref{sec:SG71} is devoted to the results in the $Immm$ phase.
The electronic spectra are described by the band structures (Fig.~\ref{fig:sup_SG71_bands}) and Fermi surfaces (Fig.~\ref{fig:sup_SG71_fermi}), all of which exhibit the same nesting mechanisms discussed in Section~\ref{sec:fermi}.
The resulting phonon instabilities are also equivalent (see Fig.~\ref{fig:sup_SG71_phonons}), and the BOES analysis (Figs.~\ref{fig:sup_SG71_BOES1} and \ref{fig:sup_SG71_BOES2}) points to the same condensation mechanism.

\begin{figure}[H]
    \centering
    \includegraphics[width=\linewidth]{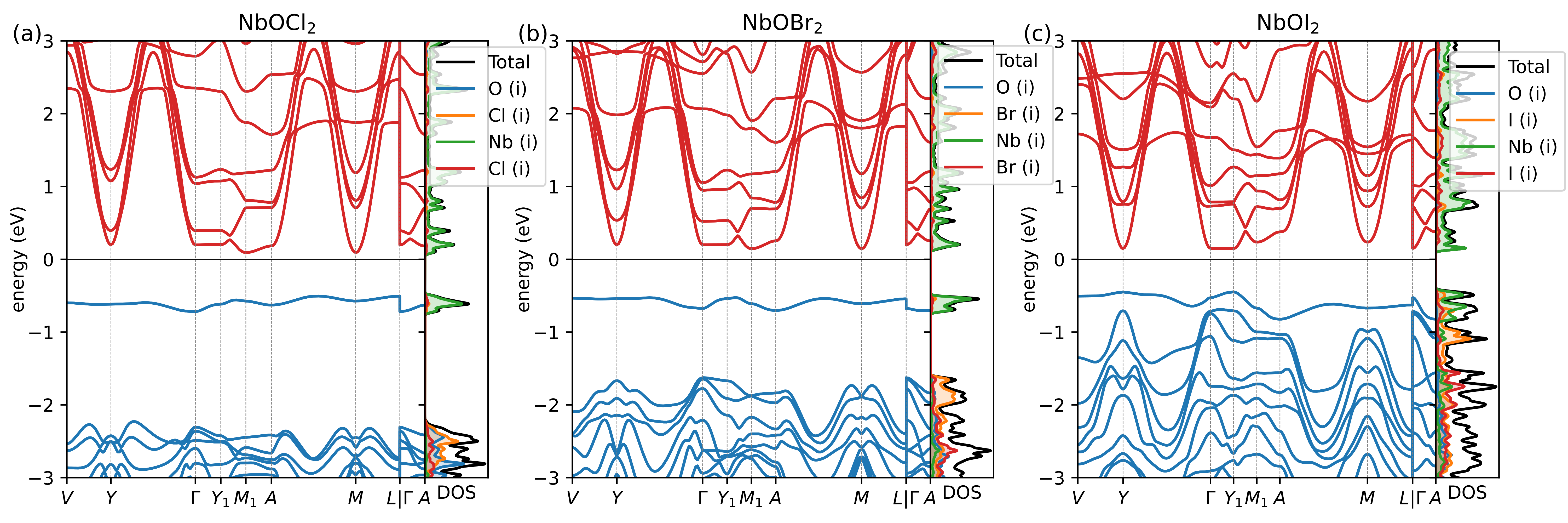}
    \caption{
        \textbf{Bandstructures for SG $\mathbf{C 2/m}$ (No.12).}
        Band structures and projected density of states for $\mathrm{NbOCl_2}$, $\mathrm{NbOBr_2}$, and $\mathrm{NbOI_2}$ are shown in panels $(a)$, $(b)$, and $(c)$, respectively.
        Labels indicate the Wyckoff position of each contribution.
        All compounds share the same flat niobium flat band just below the Fermi level.
    }
    \label{fig:sup_SG12_bands}
\end{figure}

\begin{figure}[H]
    \centering
    \includegraphics[width=\linewidth]{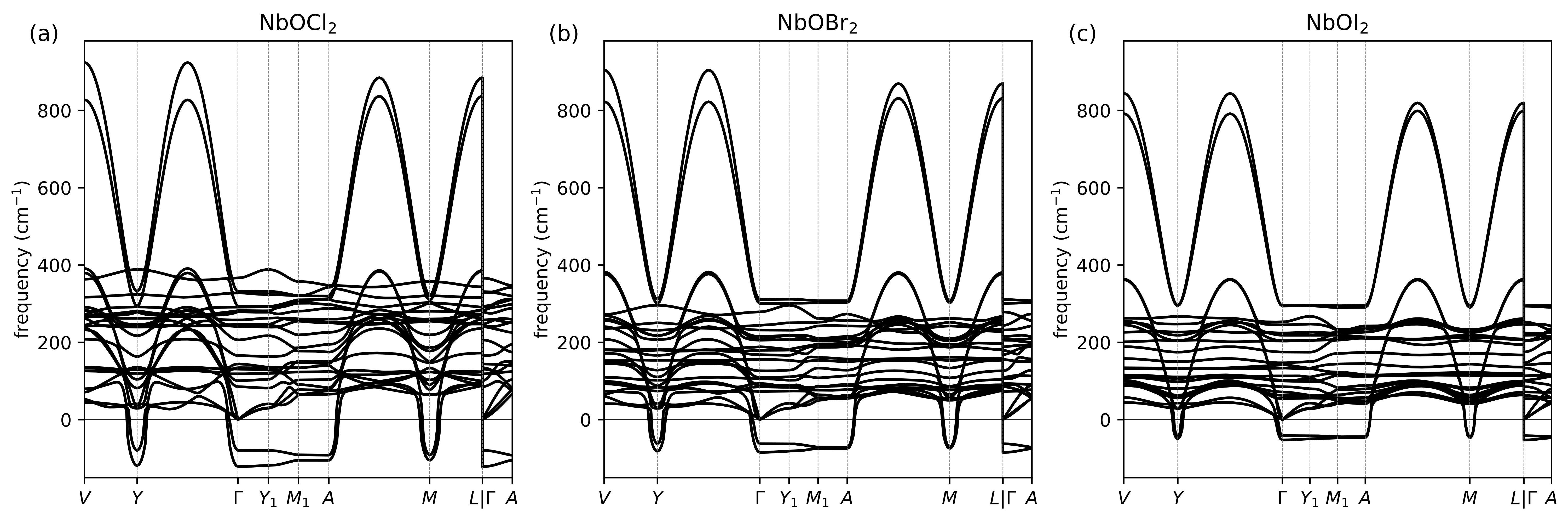}
    \caption{
        \textbf{Phonons for SG $\mathbf{C 2/m}$ (No.12).}
        Phonon spectra for $\mathrm{NbOCl_2}$, $\mathrm{NbOBr_2}$, and $\mathrm{NbOI_2}$ are shown in panels $(a)$, $(b)$, and $(c)$, respectively.
        The same instabilities are present in all compounds.
    }
    \label{fig:sup_SG12_phonons}
\end{figure}

\begin{figure}[H]
    \centering
    \includegraphics[width=\linewidth]{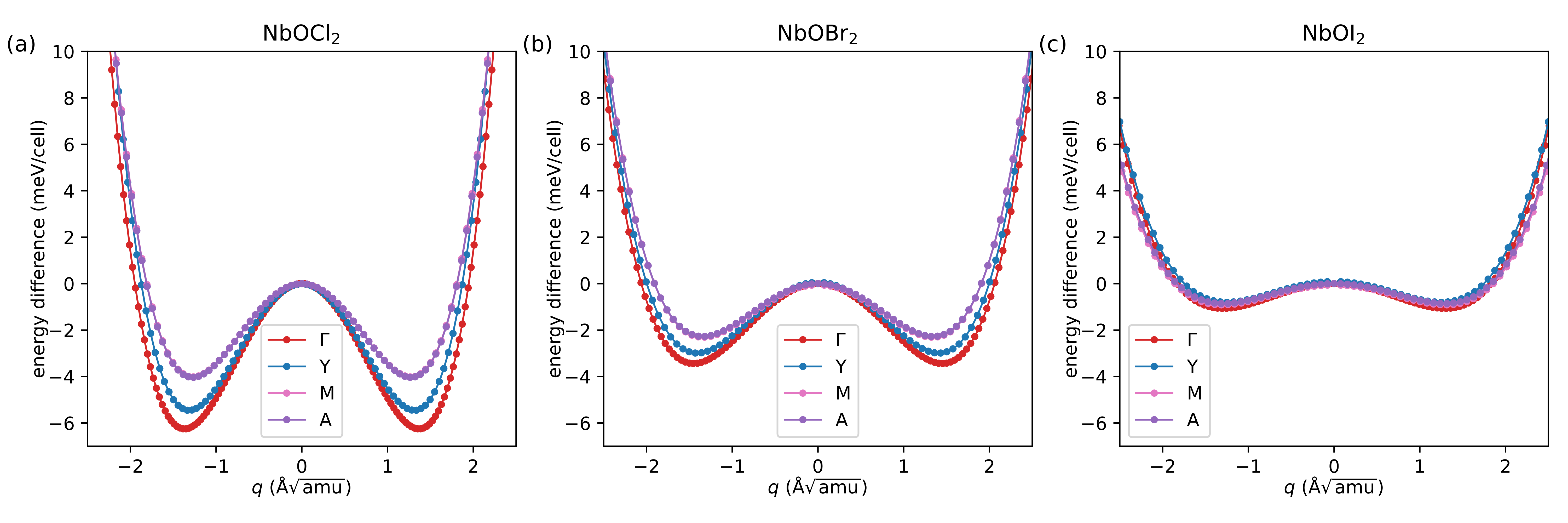}
    \caption{
        \textbf{$\mathbf{\Gamma,\ Y,\ M}$ and $\mathbf{A}$ BOES for SG $\mathbf{C2/m}$ (No.12).}
        Computed Born-Oppenheimer energy surfaces as the $C2/m$ structure is distorted parallel to the $\Gamma$, $Y$, $M$, and $A$ unstable modes for $\mathrm{NbOCl_2}$, $\mathrm{NbOBr_2}$, and $\mathrm{NbOI_2}$ are shown in panels $(a)$, $(b)$, and $(c)$, respectively.
        In all compounds, the $\Gamma$ mode leads to a lower ground energy state.
    }
    \label{fig:sup_SG12_BOES}
\end{figure}

\begin{figure}[H]
    \centering
    \includegraphics[width=\linewidth]{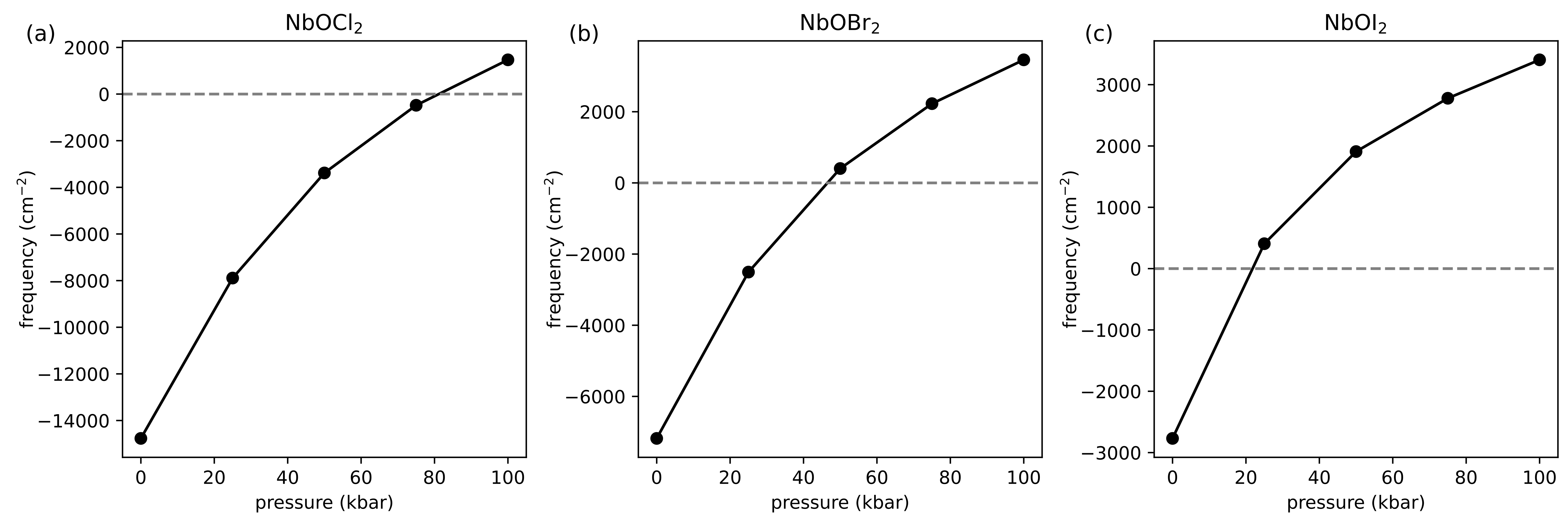}
    \caption{
        \textbf{$\mathbf{\Gamma}$ Mode as a function of pressure for SG $\mathbf{C2/m}$ (No.12).}
        The frequency evolution of the $\Gamma_1^{-}$ unstable mode as a function of pressure for $\mathrm{NbOCl_2}$, $\mathrm{NbOBr_2}$, and $\mathrm{NbOI_2}$ is shown in panels $(a)$, $(b)$, and $(c)$, respectively.
        The stabilization of the $\Gamma_1^{-}$ mode through the application of pressure demonstrates that the $C2/m$ phase can be stabilized for all compounds at approximate pressures of 81 kbar for $\mathrm{NbOCl_2}$, 46 bar for $\mathrm{NbOBr_2}$, and 21 kbar for $\mathrm{NbOI_2}$.
    }
    \label{fig:sup_SG12_pressure}
\end{figure}

In Subsection~\ref{sec:SG12}, we report the results obtained for the $C2/m$ phase.
All three compounds are insulating and feature a niobium $d$-orbital-derived flat band below the Fermi level (see Fig.~\ref{fig:sup_SG12_bands}).
Similarly, the phonon spectra shown in Fig.~\ref{fig:sup_SG12_phonons} display the same instabilities for all three compounds.
The energy minima associated with all unstable modes are compared in Fig.~\ref{fig:sup_SG12_BOES}, with the $\Gamma_1^{-}$ mode dominant in all cases.
Nonetheless, the energy difference between these minima and the intermediate $C2/m$ phase is small, suggesting that the instabilities could be stabilized by external perturbations such as pressure, or by thermal ionic fluctuations manifesting as anharmonic effects.
This is confirmed by the evolution of the $\Gamma_1^{-}$ mode with respect to pressure (Fig.~\ref{fig:sup_SG12_pressure}), which stabilizes the $C2/m$ phase at approximately 81, 46, and 21 kbar for $\mathrm{NbOCl_2}$, $\mathrm{NbOBr_2}$, and $\mathrm{NbOI_2}$, respectively.
Finally, we computed the entire phonon spectra for the three compounds at the pressures where the $\Gamma_1^{-}$ phonon was stabilized and confirmed that the spectra were stable (see Fig.~\ref{fig:sup_SG12_pressure_phonons}).

Subsection~\ref{sec:SG5} presents the results for the $C2$ phase.
All three compounds retain their insulating behavior and the niobium $d$-orbital flat band below the Fermi level observed in the $C2/m$ phase (see Fig.~\ref{fig:sup_SG5_bands}).
In this case, however, all three compounds appear dynamically stable, as indicated by the phonon spectra in Fig.~\ref{fig:sup_SG5_phonons}, which show no imaginary (negative) frequencies.

\subsection{Crystalline structures}\label{sec:structures}

\begin{table}[H]
    \begin{ruledtabular}
        \begin{tabular}{ccccc}
            \multicolumn{5}{c}{\textbf{SG $\mathbf{I m m m}$ (No. 71)}} \\ \hline
            & & $a=3.291$ \AA & $b=3.848$ \AA  &  $c=12.858$ \AA \\ \hline
            atom & Wyckoff pos. & $x$ & $y$ & $z$ \\ \hline
            Nb & $2a$ & $0$ & $0$ & $0$ \\
            O & $2d$ & $0$ & $1/2$ & $0$ \\
            Cl & $4j$ & $1/2$ & $0$ & $0.855$ \\ \hline \hline
            \\

            \multicolumn{5}{c}{\textbf{SG $\mathbf{C 2/m}$ (No. 12)}} \\ \hline
            \multicolumn{2}{c}{$\beta=105.46^{\circ}$} & $a=13.21$ \AA & $b=3.844$ \AA  &  $c=6.719$ \AA \\ \hline
            atom & Wyckoff pos. & $x$ & $y$ & $z$ \\ \hline
            Nb & $4i$ & $0$ & $0$ & $0.279$ \\
            O & $4i$ & $0.499$ & $0$ & $0.73$ \\
            Cl & $4i$ & $0.134$ & $0$ & $0.069$ \\
            Cl & $4i$ & $0.153$ & $0$ & $0.581$ \\ \hline \hline
            \\

            \multicolumn{5}{c}{\textbf{SG $\mathbf{C 2}$ (No. 5)}} \\ \hline
            \multicolumn{2}{c}{$\beta=105.55^{\circ}$} & $a=13.133$ \AA & $b=3.889$ \AA  &  $c=6.713$ \AA \\ \hline
            atom & Wyckoff pos. & $x$ & $y$ & $z$ \\ \hline
            Nb & $4c$ & $0$ & $0.976$ & $0.779$ \\
            O & $4c$ & $0.001$ & $0.509$ & $0.772$ \\
            Cl & $4c$ & $0.347$ & $0.515$ & $0.919$ \\
            Cl & $4c$ & $0.366$ & $0.521$ & $0.431$
        \end{tabular}
    \end{ruledtabular}
    \caption{\label{tab:Cl}
        Calculated lattice parameters and atomic coordinates of all the studied $\mathrm{NbOCl_2}$ phases.
        The $x,y$ and $z$ coordinates are in crystal units.
    }
\end{table}

\begin{table}[H]
    \begin{ruledtabular}
        \begin{tabular}{ccccc}
            \multicolumn{5}{c}{\textbf{SG $\mathbf{I m m m}$ (No. 71)}} \\ \hline
            & & $a=3.446$ \AA & $b=3.849$ \AA  &  $c=13.632$ \AA \\ \hline
            atom & Wyckoff pos. & $x$ & $y$ & $z$ \\ \hline
            Nb & $2a$ & $0$ & $0$ & $0$ \\
            O & $2d$ & $0$ & $1/2$ & $0$ \\
            Br & $4j$ & $1/2$ & $0$ & $0.853$ \\ \hline \hline
            \\

            \multicolumn{5}{c}{\textbf{SG $\mathbf{C 2/m}$ (No. 12)}} \\ \hline
            \multicolumn{2}{c}{$\beta=105.56^{\circ}$} & $a=14.016$ \AA & $b=3.847$ \AA  &  $c=7.06$ \AA \\ \hline
            atom & Wyckoff pos. & $x$ & $y$ & $z$ \\ \hline
            Nb & $4i$ & $0$ & $0$ & $0.284$ \\
            O & $4i$ & $0.501$ & $0$ & $0.276$ \\
            Br & $4i$ & $0.133$ & $0$ & $0.07$ \\
            Br & $4i$ & $0.155$ & $0$ & $0.584$ \\ \hline \hline
            \\

            \multicolumn{5}{c}{\textbf{SG $\mathbf{C 2}$ (No. 5)}} \\ \hline
            \multicolumn{2}{c}{$\beta=105.55^{\circ}$} & $a=13.98$ \AA & $b=3.879$ \AA  &  $c=7.053$ \AA \\ \hline
            atom & Wyckoff pos. & $x$ & $y$ & $z$ \\ \hline
            Nb & $4c$ & $0$ & $0.976$ & $0.784$ \\
            O & $4c$ & $0.001$ & $0.504$ & $0.777$ \\
            Br & $4c$ & $0.345$ & $0.51$ & $0.916$ \\
            Br & $4c$ & $0.367$ & $0.515$ & $0.43$
        \end{tabular}
    \end{ruledtabular}
    \caption{\label{tab:Br}
        Calculated lattice parameters and atomic coordinates of all the studied $\mathrm{NbOBr_2}$ phases.
        The $x,y$ and $z$ coordinates are in crystal units.
    }
\end{table}

\begin{table}[H]
    \begin{ruledtabular}
        \begin{tabular}{ccccc}
            \multicolumn{5}{c}{\textbf{SG $\mathbf{I m m m}$ (No. 71)}} \\ \hline
            & & $a=3.731$ \AA & $b=3.874$ \AA  &  $c=14.713$ \AA \\ \hline
            atom & Wyckoff pos. & $x$ & $y$ & $z$ \\ \hline
            Nb & $2a$ & $0$ & $0$ & $0$ \\
            O & $2d$ & $0$ & $1/2$ & $0$ \\
            I & $4j$ & $1/2$ & $0$ & $0.854$ \\ \hline \hline
            \\

            \multicolumn{5}{c}{\textbf{SG $\mathbf{C 2/m}$ (No. 12)}} \\ \hline
            \multicolumn{2}{c}{$\beta=105.63^{\circ}$} & $a=15.213$ \AA & $b=3.872$ \AA  &  $c=7.591$ \AA \\ \hline
            atom & Wyckoff pos. & $x$ & $y$ & $z$ \\ \hline
            Nb & $4i$ & $0.001$ & $0$ & $0.29$ \\
            O & $4i$ & $0.501$ & $0$ & $0.285$ \\
            I & $4i$ & $0.13$ & $0$ & $0.069$ \\
            I & $4i$ & $0.156$ & $0$ & $0.586$ \\ \hline \hline
            \\

            \multicolumn{5}{c}{\textbf{SG $\mathbf{C 2}$ (No. 5)}} \\ \hline
            \multicolumn{2}{c}{$\beta=105.65^{\circ}$} & $a=15.195$ \AA & $b=3.891$ \AA  &  $c=7.588$ \AA \\ \hline
            atom & Wyckoff pos. & $x$ & $y$ & $z$ \\ \hline
            Nb & $4c$ & $0.999$ & $0.98$ & $0.71$ \\
            O & $4c$ & $0.999$ & $0.502$ & $0.715$ \\
            I & $4c$ & $0.37$ & $0.51$ & $0.931$ \\
            I & $4c$ & $0.344$ & $0.507$ & $0.413$
        \end{tabular}
    \end{ruledtabular}
    \caption{\label{tab:I}
        Calculated lattice parameters and atomic coordinates of all the studied $\mathrm{NbOI_2}$ phases.
        The $x,y$ and $z$ coordinates are in crystal units.
    }
\end{table}

\subsection{SG $\mathbf{Im m m}$ (No. 71) high-symmetry phase}\label{sec:SG71}

\subsection{SG $\mathbf{C 2/m}$ (No. 12) intermediate phase}\label{sec:SG12}

\begin{figure}[H]
    \centering
    \includegraphics[width=\linewidth]{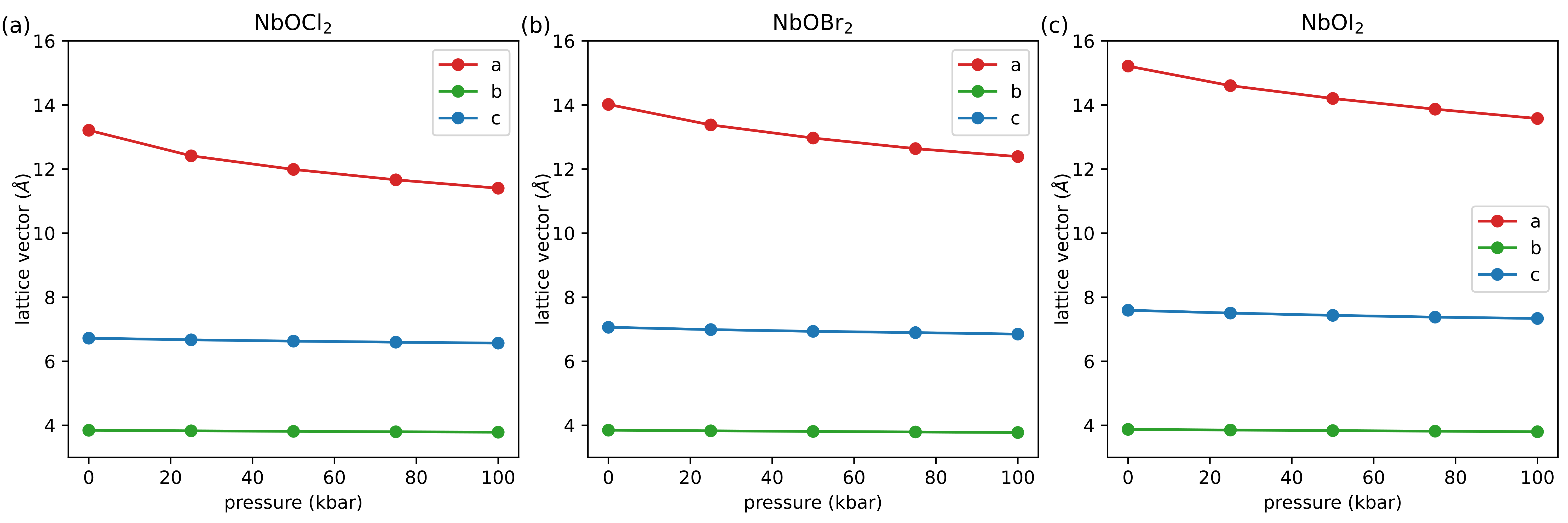}
    \caption{
        \textbf{Lattice vectors as a function of pressure for SG $\mathbf{C2/m}$ (No.12).}
        The evolution of lattice vectors as pressure is increased for $\mathrm{NbOCl_2}$, $\mathrm{NbOBr_2}$, and $\mathrm{NbOI_2}$ is shown in panels $(a)$, $(b)$, and $(c)$, respectively.
        The lattice vectors are provided in standard cell notation, as in Tables~\ref{tab:Cl}, \ref{tab:Br}, and \ref{tab:I}.
    }
    \label{fig:sup_SG12_pressure_lattice}
\end{figure}

\begin{figure}[H]
    \centering
    \includegraphics[width=\linewidth]{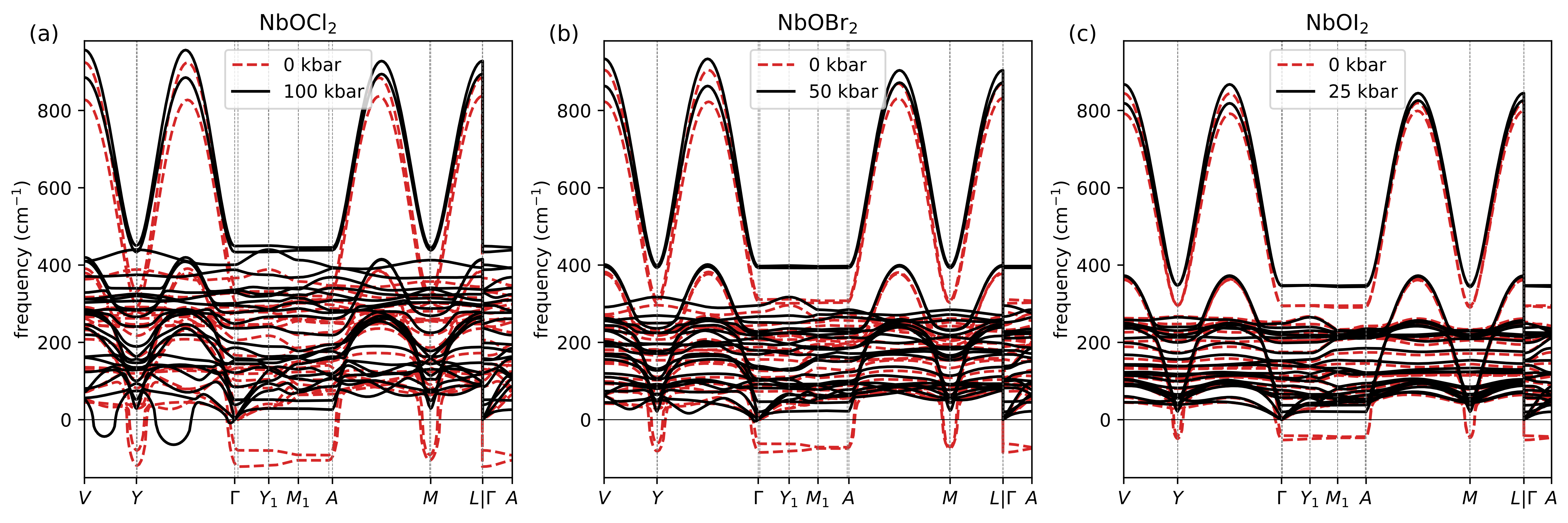}
    \caption{
        \textbf{Phonons for SG $\mathbf{C2/m}$ (No.12) under pressure.}
        As a sanity check, we computed the entire phonon spectra for $\mathrm{NbOCl_2}$, $\mathrm{NbOBr_2}$, and $\mathrm{NbOI_2}$ at the pressures where the $\Gamma_1^{-}$ phonon was stabilized for the calculations shown in Fig.~\ref{fig:sup_SG12_pressure}.
        All unstable modes are stabilized at the corresponding pressures of 100, 50, and 25 kbar.
        Note that the negative phonons appearing in panel $(a)$ for $\mathrm{NbOCl_2}$ near the $Y$ high-symmetry point are interpolation artifacts; all points explicitly within the $2\times 2\times 2$ grid show no instability.
    }
    \label{fig:sup_SG12_pressure_phonons}
\end{figure}

\subsection{SG $\mathbf{C 2}$ (No. 5) low-symmetry phase}\label{sec:SG5}

\begin{figure}[H]
    \centering
    \includegraphics[width=\linewidth]{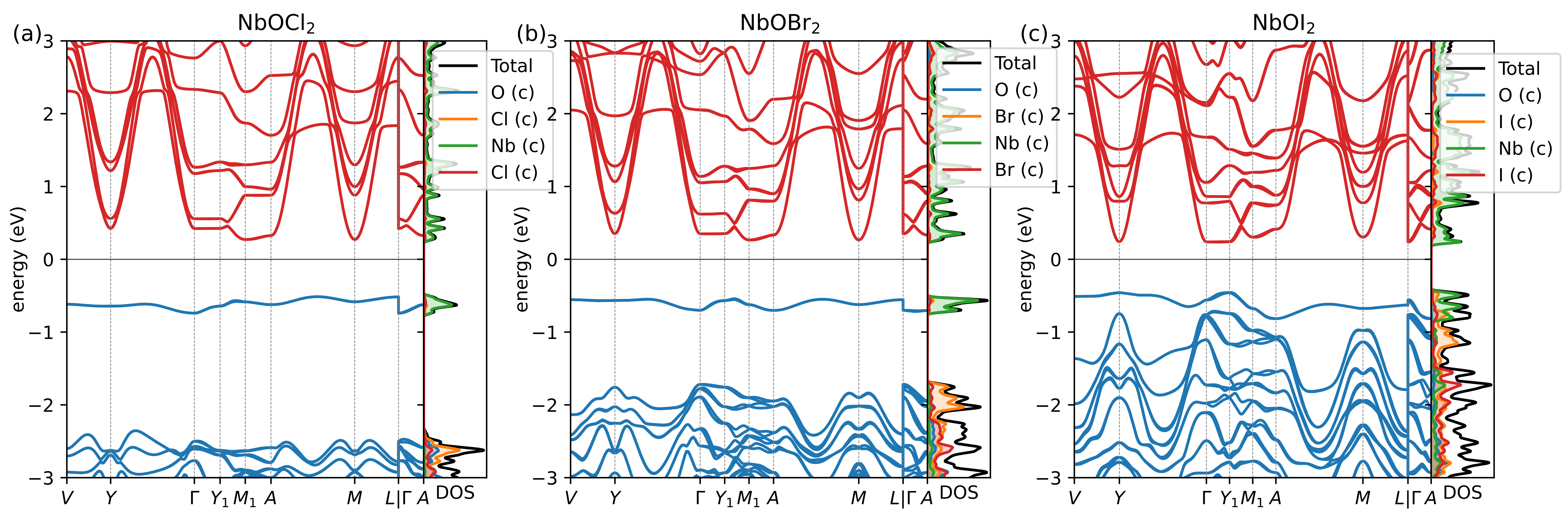}
    \caption{
        \textbf{Bandstructures for SG $\mathbf{C 2}$ (No.5).}
        Band structures and projected density of states for $\mathrm{NbOCl_2}$, $\mathrm{NbOBr_2}$, and $\mathrm{NbOI_2}$ are shown in panels $(a)$, $(b)$, and $(c)$, respectively.
        Labels indicate the Wyckoff position of each contribution.
        All compounds share the same flat niobium flat band just below the Fermi level.
    }
    \label{fig:sup_SG5_bands}
\end{figure}

\begin{figure}[H]
    \centering
    \includegraphics[width=\linewidth]{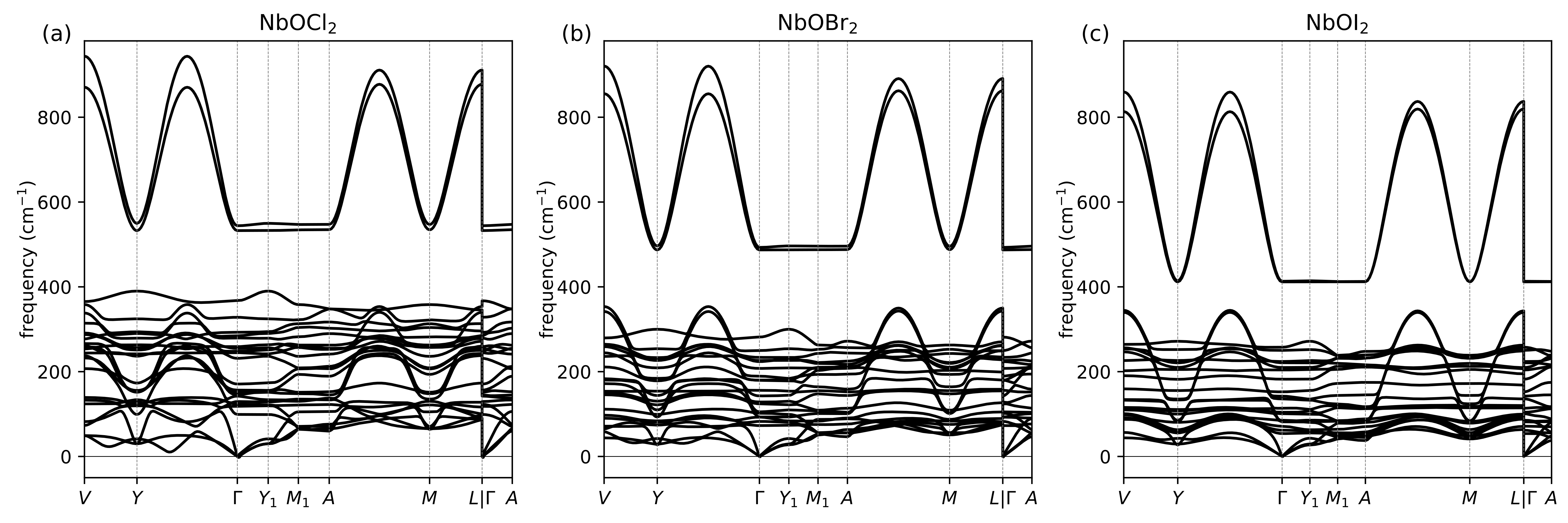}
    \caption{
        \textbf{Phonons for SG $\mathbf{C 2}$ (No.5).}
        Phonon spectra for $\mathrm{NbOCl_2}$, $\mathrm{NbOBr_2}$, and $\mathrm{NbOI_2}$ are shown in panels $(a)$, $(b)$, and $(c)$, respectively.
        The phonon spectra suggests that all compounds are stable in the $C 2$ phase.
    }
    \label{fig:sup_SG5_phonons}
\end{figure}

\clearpage %
\bibliographystyle{apsrev4-1} %
\bibliography{bibtex.bib}